\documentclass{aa}
\usepackage{tabularx}
\usepackage{amssymb}
\usepackage{amsmath}
\usepackage{txfonts}
\usepackage{balance}
\usepackage{natbib}
\bibpunct{(}{)}{;}{a}{}{,} 
\usepackage{color}
\usepackage{xspace}
\usepackage[normalem]{ulem} 
\usepackage[table]{xcolor} 
\usepackage{multirow}
\usepackage{rotating}
\usepackage{bbm}
\usepackage{cancel}
\usepackage{graphicx}
\usepackage{caption}
\usepackage{ifthen}
\usepackage{booktabs}
\usepackage{tabularx}
\usepackage{float}
\newcommand{\e}[1]{\ensuremath{\times 10^{#1}}}

\newcommand{\half}{\frac{1}{2}\xspace}
\newcommand{\St}{\text{St}\xspace}
\newcommand{\rhos}{\ensuremath{\rho_\text{s}}\xspace}
\newcommand{\rhodust}{\ensuremath{\rho_\mathrm{d}}\xspace}

\newcommand{\mf}{\ensuremath{m_\text{f}}\xspace}

\newcommand{\uf}{\ensuremath{u_\text{f}}\xspace}
\newcommand{\csound}{\ensuremath{c_\mathrm{s}}\xspace}
\newcommand{\Hp}{\ensuremath{H_\mathrm{p}}\xspace}
\newcommand{\Hg}{\ensuremath{H_\mathrm{g}}\xspace}
\newcommand{\Hd}{\ensuremath{H_\mathrm{d}}\xspace}
\newcommand{\mpr}{\ensuremath{m_\mathrm{p}}\xspace}
\newcommand{\kb}{\ensuremath{k_\mathrm{b}}\xspace}
\newcommand{\Ok}{\ensuremath{\Omega_\mathrm{k}}\xspace}

\newcommand{\Siggas}{\ensuremath{\Sigma_\mathrm{g}}\xspace}
\newcommand{\Sigdust}{\ensuremath{\Sigma_\mathrm{d}}\xspace}
\newcommand{\sighyd}{\ensuremath{\sigma_{\mathrm{H}_2}}\xspace}

\newcommand{\dx}{\ensuremath{\mathrm{d}}\xspace}

\newcommand{\aBT}{\ensuremath{a_\mathrm{BT}}\xspace}
\newcommand{\aonetwo}{\ensuremath{a_{12}}\xspace}
\newcommand{\amax}{\ensuremath{a_\mathrm{max}}\xspace}
\newcommand{\alphat}{\ensuremath{\alpha_\text{t}}\xspace}
\newcommand{\ugas}{\ensuremath{u_\text{gas}}\xspace}
\newcommand{\asett}{\ensuremath{a_\mathrm{sett}}\xspace}
\newcommand{\Rey}{\ensuremath{\mathrm{Re}}\xspace}
\makeatletter
\if@referee
    \newcommand{\com}[1]{\textcolor{red}{[#1]}}                            
    \newcommand{\highlight}[1]{\textcolor{red}{{#1\xspace}}}               
    \newcommand{\change}[2]{\sout{#1}\xspace\textcolor{red}{{#2\xspace}}}  
    \newcommand{\ccol}{{\cellcolor{red!20}}}                               
    \usepackage[nolists,noheads,nomarkers]{endfloat}                       
\else
    \newcommand{\com}[1]{}                                                 
    \newcommand{\highlight}[1]{#1\xspace}                                  
    \newcommand{\change}[2]{#2\xspace}                                     
    \newcommand{\ccol}{}                                                   
\fi
\makeatother
\usepackage[%
pdfauthor={Tilman Birnstiel},
pdftitle={Dust size distributions in coagulation/fragmentation equilibrium: Numerical solutions and analytical fits},
pdfstartview=FitH,
linkcolor=blue,
anchorcolor=blue,
citecolor=blue,
filecolor=blue,
menucolor=blue,
urlcolor=blue,
colorlinks=true]{hyperref}

\begin{document}
\title{Dust size distributions in coagulation/fragmentation equilibrium: Numerical solutions and analytical fits}
\titlerunning{Size distributions of grains in coagulation/fragmentation equilibrium}
\author{T.~Birnstiel \and C.W.~Ormel \and C.P.~Dullemond}
\authorrunning{T.~Birnstiel~et~al.}
\institute{Max-Planck-Institut f\"ur Astronomie, K\"onigstuhl 17, D-69117 Heidelberg, Germany.\\Email: birnstiel@mpia.de}
\date{\today}

\abstract
{Grains in circumstellar disks are believed to grow by mutual collisions and subsequent sticking due to surface forces. Results of many fields of research involving circumstellar disks, such as radiative transfer calculations, disk chemistry, magneto-hydrodynamic simulations largely depend on the unknown grain size distribution.}
{As detailed calculations of grain growth and fragmentation are both numerically challenging and computationally expensive, we aim to find simple recipes and analytical solutions for the grain size distribution in circumstellar disks for a scenario in which grain growth is limited by fragmentation and radial drift can be neglected.}
{We generalize previous analytical work on self-similar steady-state grain distributions. Numerical simulations are carried out to identify under which conditions the grain size distributions can be understood in terms of a combination of power-law distributions. A physically motivated fitting formula for grain size distributions is derived using our analytical predictions and numerical simulations.}
{We find good agreement between analytical results and numerical solutions of the Smoluchowski equation for simple shapes of the kernel function. The results for more complicated and realistic cases can be fitted with a physically motivated ``black box'' recipe presented in this paper. \change{}{Our results show that the shape of the dust distribution is mostly dominated by the gas surface density (not the dust-to-gas ratio), the turbulence strength and the temperature and does not obey an MRN type distribution.}}
{}

\keywords{accretion, accretion disks -- protoplanetary disks -- stars: pre-main-sequence, circumstellar matter -- planets and satellites: formation}

\maketitle

\section{Introduction}\label{sec:distri:introduction}
Dust size distributions are a fundamental ingredient for many astrophysical models in the context of circumstellar disks and planet formation: whenever dust is present, it dominates the opacity of the disk, thereby influencing the temperature and consequently also the vertical structure of the disk \citep[e.g.,][]{Dullemond:2004p390}. Small grains effectively sweep up electrons and therefore strongly affect the chemistry and the ionization fraction (also via grain surface reactions, see Vasyunin et al., in prep.) and thereby also the angular momentum transfer of the disk \citep[e.g.,][]{Wardle:1999p9890,Sano:2000p9889}.

Today, it is well established that the dust distributions in asteroid belts and debris disks are governed by a so-called ``collision cascade'' \citep[see][]{Williams:1994p9637}: larger bodies in a gas free environment exhibit high velocity collisions ($\gtrsim$ km s$^{-1}$), far beyond their critical fragmentation threshold, which lead to cratering or even complete shattering of these objects. The resulting fragments in turn suffer the same fate, thus producing ever smaller grains down to sizes below about a few micrometers where Poynting-Robertson drag removes the dust particles \citep[e.g.,][]{Wyatt:1999p10097}. The grain number density distribution in the case of such a fragmentation cascade has been derived by \citet{Dohnanyi:1969p7994} and \citet{Williams:1994p9637} and was found to follow a power-law number density distribution $n(m)\propto m^{-\alpha}$ with index $\alpha=\frac{11}{6}$ (which is equivalent to $n(a)\propto a^{-3.5}$), with very weak dependence on the mechanical parameters of the fragmentation process.
\citet{Tanaka:1996p2320}, \citet{Makino:1998p8778} and \citet{Kobayashi:2010p9774} showed that this result is exactly independent of the adopted collision model and that the resulting slope $\alpha$ is only determined by the mass-dependence of the collisional cross-section if the model of collisional outcome is self-similar \highlight{(in the context of fluid dynamics, the same result was independently obtained by \citealp{Hunt:1982p12530} and \citealp{Pushkin:2002p12243})}.
The value of $\frac{11}{6}$ agrees well with the size distributions of asteroids \citep[see][]{Dohnanyi:1969p7994} and of grains in the interstellar medium \citep[MRN distribution, see][]{Mathis:1977p789,Pollack:1985p804} and is thus widely applied, even at the gas-rich stage of circumstellar disks.

However, in protoplanetary disks gas drag damps the motions of particles. Very small particles are tied to the gas and, as a result, have a relative velocity low enough to make sticking feasible. The size distribution therefore deviates from the MRN power-law. Theoretical models of grain growth indicate that particles can grow to sizes much larger than a few $\mu$m \citep[see][]{Nakagawa:1981p4533,Weidenschilling:1980p4572,Weidenschilling:1984p4590,Weidenschilling:1997p4593,Dullemond:2005p378,Tanaka:2005p6703,Brauer:2008p215,Birnstiel:2009p7135}. Indeed, the observational evidence suggests that growth up to cm-sizes is possible \citep[e.g.,][]{Testi:2003p3390,Natta:2004p3169,Rodmann:2006p8905,Ricci:2010p9423}.

However, as particles grow, they become more loosely coupled to the gas. This results in an increase in the relative velocity between the particles, a common feature of most sources of particles' relative velocity (i.e., turbulence, radial drift, and settling motions). Therefore, we expect that the assumption of perfect sticking will break down at some point and other collisional outcomes (bouncing, erosion, catastrophic disruption) become possible \citep[see][]{Blum:2008p1920}. It is expected, then, that at a certain point growth will cease for the largest particles in the distribution.  Collisions involving these particles result in fragmentation, thus replenishing the small grains. On the other hand mutual collisions among small particles still result in coagulation.  As a result, a steady-state emerges.  In this paper, this situation is referred to as a fragmentation-coagulation equilibrium.

The situation in protoplanetary disks differs, therefore, from that in debris disks. In the latter only fragmentation operates. The mass distribution still proceeds towards a steady state but, ultimately, mass is removed from the system due to, radiation pressure or Poynting-Robertson drag.

In this paper, we consider the situation that the total mass budget in the system is conserved. For simplicity, we will ignore motions due to radial drift in this study.  This mechanism effectively removes dust particles from the disk as well as providing particles with a large relative motion.  However, the derived presence of mm-size dust particles in protoplanetary disks is somewhat at odds with the usual (laminar) prescriptions for the radial drift rate \citep{Weidenschilling:1977p865}. Recently, it was shown that mm observations of protoplanetary disks can be explained by steady-state size distributions if radial drift is inefficient \citep{Birnstiel:2010p12008}. On the other hand, if radial drift would operate as effectively as the laminar theory predicts, then the observed populations of mm-sized particles at large disk radii cannot be sustained \citep{Brauer:2007p232}. Possible solutions to reduce the drift rate include bumps in the radial pressure profile \cite[see][]{Kretke:2007p697,Brauer:2008p212,Cossins:2009p3794} or zonal flows \citep[see][]{Johansen:2009p7441}.

In this work we analytically derive steady-state distributions of grains in the presence of both coagulation and fragmentation. The analytical predictions are compared to numerical simulations and applied to grain size distributions in turbulent circumstellar disks. Both the theoretical and numerical results presented in this work are used to derive a fitting formula for steady-state grain size distributions in circumstellar disks.  

The paper is outlined as follows: in Sect.~\ref{sec:distri:theory}, we briefly summarize and then generalize previous results by \citet{Tanaka:1996p2320} and \citet{Makino:1998p8778}. In Sect.~\ref{sec:distri:simresults}, we test our theoretical predictions and their limitations by a state-of-the-art grain evolution code \citep[see][]{Birnstiel:2010p9709}. Grain size distributions in circumstellar disks are discussed in Sect.~\ref{sec:distri:diskdistris}. In Sect.~\ref{sec:distri:recipe}, we present a fitting recipe for these distributions that can easily be used in models where grain properties are important. Our findings are summarized in Sect.~\ref{sec:distri:conclusions}.

\section{Power-law solutions for a dust coagulation-fragmentation equilibrium}\label{sec:distri:theory}
In this section, we begin by summarizing some of the previous work on analytical self-similar grain size distributions on which our subsequent analysis is based. We will then extend this to include both coagulation and fragmentation processes in a common framework. Under the assumption that the relevant quantities, i.e., the collisional probability between particles, the distribution of fragments, and the size distribution, behave like power-laws, we will solve for the size distribution in coagulation-fragmentation equilibrium. For simplicity, we consider a single monomer size only of mass $m_0$. This is therefore the smallest mass in the distribution. 

Another key assumption of our analytical model is that we assume the existence of a sharp threshold mass $\mf$, above which collisions always result in fragmentation and below which collision always result in coagulation. As explained above, the physical motivation for this choice is that relative velocities increase with mass. This also means that in our theoretical model we neglect collision outcomes like bouncing or erosion \highlight{(erosion is included in the simulations)}. \change{}{Even though, small particles will in reality cause cratering/growth instead of complete fragmentation of the larger particle, we find that this assumption is often justified because fragmenting similar-sized collisions prevent any growth beyond \mf. Thus, collisions with much smaller particles ($m<<\mf$) do not have an important influence on the maximum size, however, they can significantly change the amount of small particles in the stationary distribution (see Sect.~\ref{sec:distri:recipe}).}
\change{}{We further assume a constant porosity of the particles, which relates the mass and size according to
\begin{equation}
m = \frac{4\,\pi}{3}\,\rhos\,a^3,
\end{equation}
where $\rhos$ is the internal density of a dust aggregate.
}

\begin{figure}[thb]
  \begin{center}
    \includegraphics[width=\columnwidth]{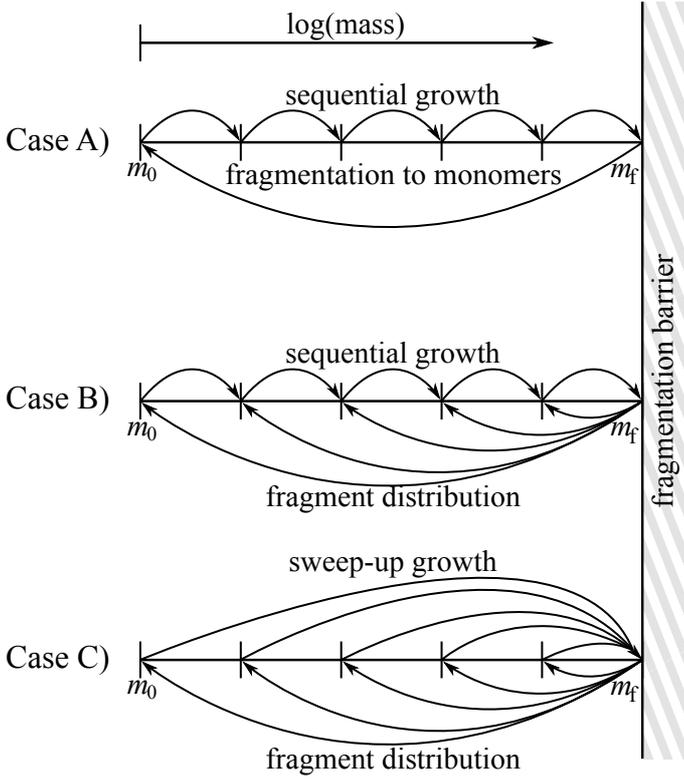}
    \caption{Illustration of the three different regimes for which analytical solutions have been derived. Case A represents the growth cascade discussed in Sect.~\ref{sec:distri:theory_coag}, case B the intermediate regime (see Sect.~\ref{sec:distri:theory_intermediate}) and case C the fragmentation dominated regime which is discussed in Sect.~\ref{sec:distri:theory_fragment}. Particles are shattered once they reach the fragmentation barrier \mf since collision velocities for particles $>\mf$ exceed the fragmentation threshold velocity.}
    \label{fig:distri:sketch}
  \end{center}
\end{figure}

In our analysis, we will identify three different regimes, which are symbolically illustrated in Fig.~\ref{fig:distri:sketch}:
\begin{itemize}
\item 
Regime A represents a case where grains grow sequentially (i.e. hierarchically) by collisions with similar sized grains until they reach an upper size limit and fragment back to the smallest sizes. The emerging power-law slope of the size distribution depends only on the shape of the collisional kernel.
\item 
Regime B is similar to regime A, however in this case the fragmented mass is redistributed over a range of sizes and, thus, influencing the out-coming distribution.
\item 
In Regime C, the upper end of the distribution dominates grain growth at all sizes. Smaller particles are swept up by the upper end of the distribution and are replenished mostly by redistributed fragments of the largest particles. The resulting distribution function depends strongly on how the fragmented mass is distributed after a disruptive collision.
\end{itemize}
For each of these regimes, we will derive the parameter ranges for which they apply and the slopes of the resulting grain size distribution.  

\subsection{The growth cascade}\label{sec:distri:theory_coag}
The fundamental quantity that governs the time-evolution of the dust size distribution is the collision kernel $C_{m_1,m_2}$. It is defined such that
\begin{equation}
C_{m_1,m_2}\cdot n(m_1)\cdot n(m_2)\, \dx{m_1} \dx{m_2}
\label{eq:distri:collisions1}
\end{equation}
gives the number of collisions per unit time per unit volume between particles in mass interval $[m_1, m_1+\dx{m_1}]$ and $[m_2,m_2+\dx{m_2}]$, where $n(m)$ is the number density distribution. Once specified it determines the collisional evolution of the system. In the case that the number density $n(m)$ does not depend on position, $C_{m_1,m_2}$ is simply the product of the collision cross section and the relative velocity of two particles with the masses $m_1$ and $m_2$. 

We use the same Ansatz as \citet{Tanaka:1996p2320}, assuming that the collision kernel is given in the self-similar form
\begin{equation}
  C_{m_1,m_2} = m_1^\nu \: h\left(\frac{m_2}{m_1}\right).
  \label{eq:distri:kernel_form}
\end{equation}
Here, $h$ is any function which depends only on the masses through the ratio of $m_2/m_1$. \change{}{By definition, the kernel $C_{m_1,m_2}$ has to be symmetric, therefore, Eq.~\ref{eq:distri:kernel_form} implies that $h(m_1,m_2)$ is not symmetric (see Eq.~\ref{eq:distri:gamma_makino}).} $\nu$ is called the index of the kernel or the degree of homogeneity.
As we will see in the following, $\nu$ is one of the most important parameters determining the resulting size distribution. Different physical environments are represented by different values of $\nu$. Examples include the constant kernel (i.e., $\nu=0$, mass independent), the geometrical kernel (i.e., $\nu=2/3$, velocity independent) or the linear kernel (i.e., $\nu=1$, as for grains suspended in turbulent gas).
 
It is further assumed that the number density distribution of dust particles follows a power-law,
\begin{equation}
  n(m) = A \cdot  m^{-\alpha}.
  \label{eq:distri:massdistribution}
\end{equation}

The time evolution of the mass distribution obeys the equation of mass conservation,
\begin{equation}
  \frac{\partial m n(m)}{\partial t} + \frac{\partial F(m)}{\partial m} = 0,
  \label{eq:distri:massconservation}
\end{equation}
where the flux $F(m)$ does not represent a flux in a typical continuous way since coagulation is non-local in mass space (each mass can interact with each other mass) but rather an integration of all growth processes which produce a particle with mass greater than $m$ out of a particle that was smaller than $m$ (i.e. collisions of $m_1<m$ with any other mass $m_2$ such that $m_1+m_2>m$). The flux in the case of pure coagulation was derived by \citet{Tanaka:1996p2320}
\begin{equation}
  F(m)   =  \int_{m_0}^m \mathrm{d}m_1 \int_{m-m_1}^{\mf} \mathrm{d}m_2 \, m_1 \, C_{m_1,m_2} \, n(m_1) \, n(m_2),
  \label{eq:distri:tanaka_flux_1}
\end{equation}
where we changed the lower bound of the integration over $m_1$ to start from the mass of monomers $m_0$ (instead of 0) and the upper bound of the integral over $m_2$ to a finite upper end $\mf$ (instead of infinity), compared to the definition used by \citet{Tanaka:1996p2320}.

Substituting the definitions of above and using the dimensionless variables $x_1 = m_1/m$, $x_2 = m_2/m$, $x_0 = m_0/m$, and $x_\mathrm{f} = \mf/m$ one obtains
\begin{equation}
  F(m)   =  m^{\nu-2\alpha+3} \underbrace{\int_{x_0}^1 \mathrm{d}x_1\int_{1-x_1}^{x_\mathrm{f}} \mathrm{d}x_2 x_1^{\nu+1-\alpha} x_2^{-\alpha} h\left(\frac{x_2}{x_1}\right)}_\text{:= K},
  \label{eq:distri:tanaka_flux_2}
\end{equation}
where $K$ \change{integrates to a constant}{approaches a constant value in the limit of $m \gg m_0$ and $m \ll \mf$}.

Postulation of a steady state (i.e. setting the time derivative in Eq.~\ref{eq:distri:massconservation} to zero), leads to the condition
\begin{equation}
  F(m) \propto m^{\nu-2\alpha+3} = const.,
\end{equation}
from which it follows that the slope of the distribution is
\begin{equation}
  \alpha = \frac{\nu+3}{2}.
  \label{eq:distri:coag_cascade}
\end{equation}
This result was already derived for the case of fragmentation by \citet{Tanaka:1996p2320} and \citet{Dohnanyi:1969p7994} and for the coagulation by \citet{Klett:1975p3936} and \citet{Camacho:2001p3816}.
The physical interpretation of this is a ``reversed'' fragmentation cascade: instead of a resupply of large particles which produces ever smaller bodies, this represents a constant resupply of monomers which produce ever larger grains (cf. case A in Fig.~\ref{fig:distri:sketch}).

\subsection{Coagulation fragmentation equilibrium}\label{sec:distri:theory_intermediate}
As \citet{Tanaka:1996p2320} and \citet{Makino:1998p8778} pointed out, the previous result is independent of the model of collisional outcomes as long as this model is self-similar (Eq.~\ref{eq:distri:kernel_form}). However this is no longer the case if we consider both coagulation and fragmentation processes happening at the same time.

We will now consider the case with a constant resupply of matter, due to the particles that fragment above \mf.  We assume these fragments obey a power-law mass distribution and are produced at a rate
\begin{equation}
  \dot n_\mathrm{f}(m) = N \cdot m^{-\xi},
  \label{eq:distri:frag_powerlaw}
\end{equation}
where $\xi$ reflects the shape of the fragment distribution and $N$ is a constant.

If there is a constant flux of particles $F(m)$ as defined above, then the flux of fragmenting particles (i.e., the flux produced by particles that are growing over the fragmentation threshold) is given by
\begin{equation}
  F(\mf) =  K\cdot \mf^{\nu-2\alpha+3}
  \label{eq:distri:flux_mf}
\end{equation}
where $K$ is the integral defined in Eq.~\ref{eq:distri:tanaka_flux_2}.

The resulting (downward) flux of fragments $F_\text{f}(m)$ can then be derived by inserting Eq.~\ref{eq:distri:frag_powerlaw} into the equation of mass conservation (Eq.~\ref{eq:distri:massconservation}),
\begin{equation}
  \frac{\partial F_\text{f}(m)}{\partial m} = - m \cdot \dot n_\text{f}.
\end{equation}
Integration from monomer size $m_0$ to $m$ yields
\begin{equation}
  F_\text{f}(m) = - N \cdot \frac{1}{2-\xi} \cdot \left( m^{2-\xi} - m_0^{2-\xi} \right).
  \label{eq:distri:flux_downward}
\end{equation}
The normalization factor $N$ can be determined from the equilibrium condition that the net flux vanishes,
\begin{equation}
  F_\text{f}(\mf) = - F(\mf),
\end{equation}
and was found to be
\begin{equation}
N = \left(2-\xi\right) \, K \, \frac{\mf^{\nu-2\alpha+3}}{\mf^{2-\xi}-m_0^{2-\xi}}.
\label{eq:distri:normalization}
\end{equation}
In Eq.~\ref{eq:distri:flux_downward}, we can distinguish two cases:
\begin{itemize}
  \item If $\xi > 2$, the contribution of $m_0$ dominates the term in brackets. This means that
  most of the fragment mass is redistributed to monomer sizes and the situation is the same
  as in the pure coagulation case (cf. Case A in Fig.~\ref{fig:distri:sketch}). The steady-state condition
  $F(m)+F_\text{f}(m)=0$ (i.e., the net flux is zero) yields that
  \begin{equation}
    K\cdot m^{\nu-2\alpha+3}= - N \cdot \frac{1}{2-\xi} \cdot m_0^{2-\xi},
  \end{equation}
  is constant, which leads to the same result as Eq.~\ref{eq:distri:coag_cascade}. Intuitively, this is clear since the majority of the redistributed mass ends up at $m \sim m_0$.
  \item If $\xi<2$, the $m$-dependence dominates the term in brackets in Eq.~\ref{eq:distri:flux_downward} and postulation of a steady-state,
  \begin{equation}
    K\cdot m^{\nu-2\alpha+3} \simeq N \cdot \frac{1}{2-\xi} \cdot m^{2-\xi},
  \end{equation}
  leads to an exponent
  \begin{equation}
    \alpha = \frac{\nu+\xi+1}{2},
    \label{eq:distri:intermediate_slope}
  \end{equation}
  less than Eq.~\ref{eq:distri:coag_cascade}.
  In this case, the slope of the fragment distribution matters. This scenario is represented as case B in Fig.~\ref{fig:distri:sketch}. 
\end{itemize}

\subsection{Fragment dominated regime}\label{sec:distri:theory_fragment}
The result obtained in the previous section may seem to be quite general. However, it does not hold for low $\xi$-values as we will show in the following.

In our case, the integrals do not diverge due to the finite integration bounds. However \citet{Makino:1998p8778} used $0$ and $\infty$ as lower and upper bounds for the integration and thus needed to investigate the convergence of the integral. They derived the following conditions for convergence:
\begin{align}
  \nu-\gamma-\alpha+1 &< 0 \label{eq:distri:makino_convergence1} \\
  \gamma - \alpha + 2 &> 0 \label{eq:distri:makino_convergence2}. 
\end{align}
where $\gamma$ gives the dependence of $m_2/m_1$ in the $h$-function of Eq.~\ref{eq:distri:gamma_makino} \citep[see][]{Makino:1998p8778}:
\begin{equation}
  h \left( \frac{m_2}{m_1} \right) = h_0 \cdot \left\lbrace
  \begin{array}{ll}
\left( \frac{m_2}{m_1} \right) ^{\gamma}      & \text{for }\frac{m_2}{m_1}\ll 1\\
\\
\left( \frac{m_2}{m_1} \right) ^{\nu-\gamma}  & \text{for }\frac{m_2}{m_1}\gg 1
\end{array}\right.
\label{eq:distri:gamma_makino}
\end{equation}

The first condition (Eq.~\ref{eq:distri:makino_convergence1}) considers the divergence towards the upper masses, whereas the second condition pertains the lower masses.  We assume that Eq.~\ref{eq:distri:makino_convergence2} is satisfied and consider the case of decreasing $\xi$ for $\alpha$ given by Eq.~\ref{eq:distri:intermediate_slope} which results in a steeper size distribution (where the mass is concentrated close to the upper end of the distribution). We see that for $\xi<1+\nu-2\gamma$, Eq.~\ref{eq:distri:makino_convergence1} is no longer fulfilled. The behavior of the flux integral changes qualitatively: growth is no longer hierarchical, but it becomes dominated by contributions by the upper end of the integration bounds.

Physically, this means that the total number of collisions of any grain is determined by the largest grains in the upper end of the distribution. Hence, smaller sized particles are predominantly refilled by fragmentation events of larger bodies instead of coagulation events from smaller bodies and they are  predominantly removed by coagulation events with big particles (near the threshold \mf) instead of similar-sized particles.  This corresponds to Case C in Fig.~\ref{fig:distri:sketch}.

To determine the resulting power-law distribution, we again focus on Eq.~\ref{eq:distri:tanaka_flux_1}. The double integral of the flux $F(m)$ can now be split into three separate integrals,
\begin{align}
F(m) &=& \int_{m_0}^{\frac{m}{2}} \dx{m_1} \int_{m-m_1}^{\mf}  \dx{m_2} \, m_1 \, C_{m_1,m_2} \, n(m_1) \, n(m_2)\nonumber\\
& &+\int_{\frac{m}{2}}^{m}   \dx{m_1} \int_{m-m_1}^{m_1}  \dx{m_2} \, m_1 \, C_{m_1,m_2} \, n(m_1) \, n(m_2)\label{eq:distri:split_integral}\\
& &+\int_{\frac{m}{2}}^{m}   \dx{m_1} \int_{m_1}^{\mf}  \dx{m_2} \, m_1 \, C_{m_1,m_2} \, n(m_1) \, n(m_2)\nonumber
\end{align}
according to whether $m_2$ is larger or smaller than $m_1$.

It can be derived (see Appendix~\ref{sec:distri:appendix}) that if the condition above (Eq.~\ref{eq:distri:makino_convergence1}) is violated and if $\mf\gg m$, then the first and the third integral in Eq.~\ref{eq:distri:split_integral} dominate the flux due to the integration until \mf. In this cases, the flux $F(m)$ is proportional to $m^{2+\gamma-\alpha}$.

A stationary state in the presence of fragmentation (Eq.~\ref{eq:distri:flux_downward}) is reached if the fluxes cancel out, which leads to the condition
\begin{equation}
  \alpha = \xi + \gamma.
  \label{eq:distri:fragment_regime}
\end{equation}
This is the sweep-up regime where small particles are cleaned out by big ones (cf. Case C in Fig.~\ref{fig:distri:sketch}). 
\subsection{Summary of the regimes}\label{sec:distri:theory_summary}

Summarizing these findings, we find that the resulting distribution is described by three scenarios (depicted in Fig.~\ref{fig:distri:sketch}), depending on the slope of the fragment distribution:
\begin{eqnarray}
\begin{array}{lll} 
  \text{\small Case A (growth cascade):}      &\xi>2               & \alpha = \frac{\nu+3}{2}    \\
  \\
  \text{\small Case B (intermediate regime):} &\nu-2\gamma+1<\xi<2 & \alpha = \frac{\nu+\xi+1}{2} \\
  \\
  \text{\small Case C (fragment dominated):}  &\xi < \nu-2\gamma+1 & \alpha = \xi+\gamma
\end{array}
\label{eq:distri:nu_regimes}  
\end{eqnarray} 

\section{Simulation results interpreted}\label{sec:distri:simresults}
In this section, we will test the analytical predictions of the previous section by a coagulation/fragmentation code (\citealp{Birnstiel:2010p9709}, see also \citealp{Brauer:2008p215}). The code solves for the time evolution of the grain size distribution using an implicit integration scheme. This enables us to find the steady-state distribution by using large time steps. In this way, the time evolution is not resolved, but the steady-state distribution is reliably and very quickly derived.

We start out with the simplest case of a constant kernel and then -- step by step -- approach a more realistic  scenario (in the context of a protoplanetary disk). In Sect.~\ref{sec:distri:diskdistris}, we will then consider a kernel taking into account relative velocities of Brownian motion and turbulent velocities and also a fragmentation probability which depends on the masses and the relative velocities of the colliding particles.

\change{}{The following results are only steady-state solutions, whether or not this state is reached depends on several conditions. Firstly, particles need to fragment. If there is no upper boundary for growth, it will proceed unlimited and a steady state will never be reached. Secondly, radial motion needs to be slow enough to allow for a steady state. If quantities like the surface density or the temperature vary smoothly between neighbouring regions in the disk, the steady state solutions will also be similar and radial transport will happen on top of a steady-state grain distribution. However if radial drift is acting strongly on  particles of the distribution, the steady state will not be reached. Thirdly, a distribution of initially sub-$\mu$m sized grains will need some time to get into an equilibrium state. This time scale can be as small as $<$ 1000 years inside of a few AU, while it can be of the order of a million years at 100~AU. We provide a rough estimate of this time scale in Appendix~\ref{sec:distri:timescale}.}

\subsection{Constant kernel}\label{sec:distri:simresults_constant}
In the following section, we consider the case of a constant kernel and will include fragmentation above particle sizes of 1~mm because this represents an instructive test case.

\begin{figure}[thb]
  \begin{center}
\includegraphics[width=\columnwidth]{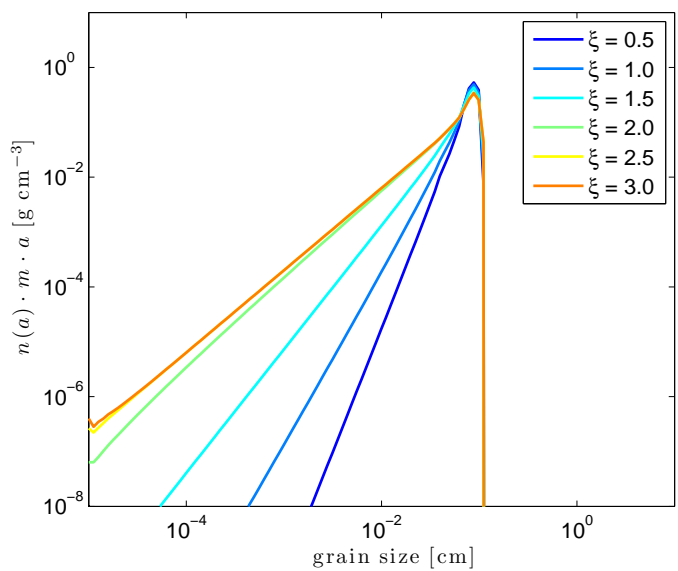}
\caption{Grain size distributions\protect\footnotemark{} for a constant kernel (i.e., $\nu=0$) and different distributions of fragments. The peak towards the upper end of the distribution is due to the fragmentation barrier (explanation in the text). The slope of the mass distribution corresponds to $6-3\alpha$.}
\label{fig:distri:spectra}
\end{center}
\end{figure}
\footnotetext{It should be noted that in this paper we will plot the distributions typically in terms of $n(a)\cdot m\cdot a$ which is proportional to the distribution of mass. The advantage of plotting it this way instead of plotting $n(a)$ is the following: when $n(m)$ follows a power-law $m^{-\alpha}$, then the grain size distribution $n(a)$ describes a power-law with exponent $2-3\alpha$ and the mass distribution exponent is $6-3\alpha$. For typical values of $\alpha$, this is less steep and differences between a predicted and a real distribution are more prominent.}

We iteratively solve for the steady-state size distribution between coagulation and fragmentation. The outcome of these simulations for a constant kernel (i.e., $\nu=0$) are power-law distributions where the slope of the distribution depends on the fragmentation law (the slope $\xi$, see Eq.~\ref{eq:distri:frag_powerlaw}). 
Figure~\ref{fig:distri:spectra} shows the corresponding size distributions for some of the different fragment distributions: the steepest distribution corresponds to the case of $\xi=0.5$. For larger $\xi$-values, the slope of the mass distribution flattens. In all cases, a bump develops towards the upper end of the distributions. The reason for this \change{}{``pile-up''} is the following: grains typically grow mostly through collisions with similar-sized or larger particles. Since the distribution is truncated at the upper end (defined as \amax, see also Eq.~\ref{eq:distri:a_max}), particles close to the upper end lack larger collision partners, the growth rate at these sizes would decrease if the distribution keeps its power-law nature. This, in turn means that the flux could not be constant below \amax. To keep a steady state, the number of particles at that point has to increase in order to replace the missing collision partners at larger sizes.   

\begin{figure}[htp]
  \begin{center}
\includegraphics[width=\columnwidth]{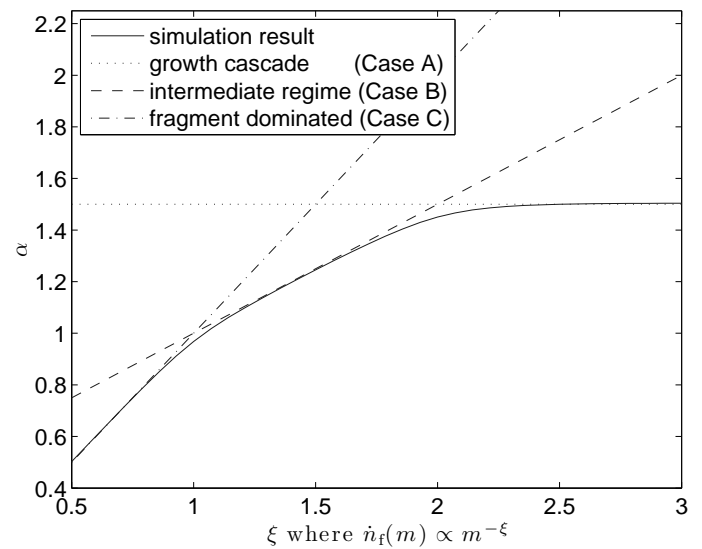}
\caption{Exponent of grain size distributions for a constant kernel (i.e. $\nu=0$) and different distributions of fragments (solid line) and the analytic solution for the growth cascade (Case A, dotted line), the intermediate regime (Case B, dashed line), and the fragment dominated regime (Case C, dash-dotted line).}
\label{fig:distri:exponent}
\end{center}
\end{figure}

Figure~\ref{fig:distri:exponent} shows how the slope of the resulting distribution depends on the fragmentation slope $\xi$, where the three previously discussed regimes can be identified:
\begin{itemize}
  \item
  Case A, growth cascade: as predicted, this scenario holds for values of $\xi \gtrsim 2$ where most of the fragmenting mass is redistributed to fragments. This case corresponds to a ``reversed'' collisional cascade (just the direction in mass space is reversed as collisions mostly lead to growth instead of shattering). 
  
  \item
  Case B, the intermediate regime: when the fragment mass is more or less equally distributed over all sizes, mass gain by redistribution of fragments and the mass loss due to coagulation have to cancel each other.
  
  \item
  Case C: most of the fragmented mass remains at large sizes. Therefore the mass distribution is dominated by the largest particles. In other words, growth is not hierarchical anymore. In this test case $\gamma=0$, and from Eq.~\ref{eq:distri:makino_convergence1} it follows that the transition between the intermediate and fragment-dominated regime lies at $\xi = 1$, which can be seen in Fig.~\ref{fig:distri:exponent}.
\end{itemize}
The measured slopes of the size distribution for $m\ll\mf$ are in excellent agreement with the model outlined in Sect.~\ref{sec:distri:theory}.
\subsection{Including settling effects}\label{sec:distri:simresults_settling}
The simplest addition to this model is settling: as grains become larger, they start to settle towards the mid-plane. However, turbulent mixing counteracts this systematic motion. The vertical distribution of dust in a settling-mixing equilibrium can (close to the mid-plane) be estimated by a Gaussian distribution with a size-dependent dust scale height \Hd.
Smaller particles are well enough coupled to the gas to have the same scale height as the gas, \Hg, while larger particles decouple and their scale height is decreasing with grain size,
\begin{equation}
\frac{H_\mathrm{d}}{\Hg} = \sqrt{\frac{\alphat}{\St}},\qquad\mathrm{for }\qquad \St>\alphat
\label{eq:distri:h_dust}
\end{equation}
\citep[see e.g.,][]{Dubrulle:1995p300,Schrapler:2004p2394,Youdin:2007p2021} where the Stokes number \St is a dimensionless quantity which describes the dynamic properties of a suspended particle. Very small particles have a small Stokes number and are therefore well coupled to the gas. Particles which have different properties (e.g., size or porosity) but the same \St behave aerodynamically the same. In our prescription of turbulence, \St is defined as the product of the particles stopping time $\tau_\mathrm{st}$ and the orbital frequency $\Omega_\mathrm{k}$. We focus on the Epstein regime where the Stokes number can be approximated by
\begin{equation}
  \St = \Ok \cdot \tau_\mathrm{st} \simeq \frac{a \, \rhos}{\Siggas} \frac{\pi}{2}
  \label{eq:distri:st}
\end{equation}
with \Siggas being the gas surface density and \rhos being the internal density of the particles which relates mass and size via $m=4\pi/3\rhos a^3$,

Settling starts to play a role as soon as the Stokes number becomes larger than the turbulence parameter $\alphat$ which can be related to a certain size
\begin{equation}
  \asett = \frac{2\alphat \, \Siggas}{\pi \rhos}.
  \label{eq:distri:asett}
\end{equation}
Eq.~\ref{eq:distri:asett} only holds within about one gas pressure scale height because the dust vertical structure higher up in the disk deviates from the a Gaussian profile \citep[see also][]{Dubrulle:1995p300,Schrapler:2004p2394,Dullemond:2004p390}.

The mass-dependent dust scale height causes the number density distribution $n(m)$ to depend on the vertical height, $z$. In disk-like configurations, it is therefore customary to consider the column density,
\begin{equation}
N(m) = \int_{-\infty}^{\infty} n(m,z) \,\dx{z}.
\end{equation}
Similar to Eq.~\ref{eq:distri:collisions1}, we can write the vertically integrated number of collisions as
\begin{equation}
\tilde{C}_{m_1,m_2}\cdot N(m_1)\cdot N(m_2) \,\dx{m_1}\, \dx{m_2},
\label{eq:distri:collsions2}
\end{equation}
which gives the total number of collisions that take place over the entire column of the disks.

The dependence of the collisional probability on scale height, is now reflected in the modified kernel $\tilde{C}_{m_1,m_2}$ \citep[see][Appendix A for derivation]{Birnstiel:2010p9709}:
\begin{equation}
  \tilde{C}_{m_1,m_2} = \frac{C_{m_1,m_2}}{\sqrt{2\pi \left(H_1^2+H_2^2\right)}}.
  \label{eq:distri:cmod_1}
\end{equation}

The point to realize here is that, due to the symmetry between Eq.~\ref{eq:distri:collisions1} and Eq.~\ref{eq:distri:collsions2}, the analysis in Sect.~\ref{sec:distri:theory} holds also for disk-like configuration, if the kernel is now replaced by $\tilde{C}_{m_1,m_2}$. The resulting exponent $\alpha$ then concerns the column density dependence ($N(m)\propto m^{-\alpha}$).

If we consider the case that $\St>\alphat$ and substitute Eq.~\ref{eq:distri:h_dust} and Eq.~\ref{eq:distri:st} into Eq.~\ref{eq:distri:cmod_1}, we find that 
\begin{equation}
  \begin{split}
\tilde{C}_{m_1,m_2} &= \frac{C_{m_1,m_2}}{\sqrt{2\pi \left(H_1^2+H_2^2\right)}}\\
&= C_{m_1,m_2}\cdot H_1^{-1} \cdot \left(1+\frac{H_2^2}{H_1^2}\right)^{-\frac{1}{2}}\\
&= C_{m_1,m_2}\cdot m_1^{1/6} \cdot h\left(\frac{m_2}{m_1}\right),\\
\end{split}
\label{eq:distri:settling_c}
\end{equation}
has an index $\nu = 1/6$ for grain sizes larger than $a_\mathrm{sett}$ and $\nu=0$ otherwise (it should be noted that $H_1$ and $H_2$ are the dust scale heights whereas $h(m_2/m_1)$ represents the function defined in Eq.~\ref{eq:distri:kernel_form}).

\begin{figure}[htp]
  \centering
  \includegraphics[width=\columnwidth]{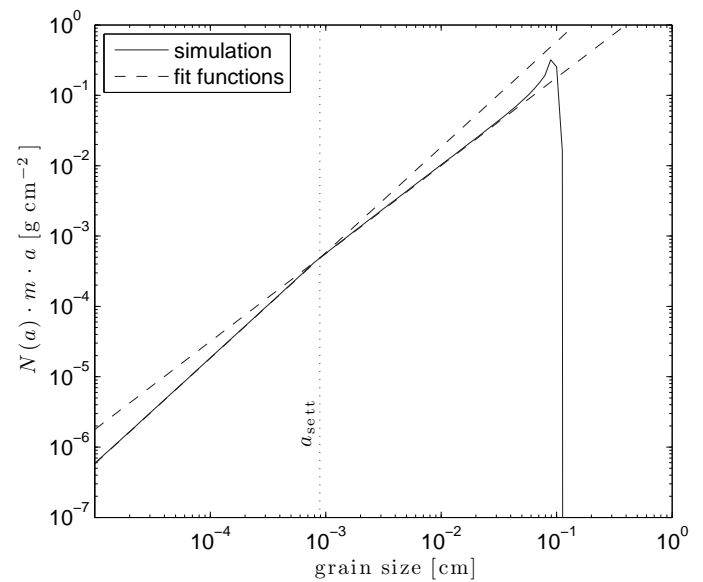}
  \caption{Simulation result (solid line) and a ``two power-law fit'' to the vertically integrated dust distribution for a constant kernel with settling included. The dotted, vertical line denotes the grain size above which grains are affected by settling.}
  \label{fig:distri:settling_fit}
\end{figure}

The theory described in Sect.~\ref{sec:distri:theory} is strictly speaking only valid for a constant $\nu$-index, but if this index is constant over a significant range of masses, then the local slope of the distribution will still adapt to this index. In the case of settling, we can therefore describe the distribution with two power-laws as can be seen in Fig.~\ref{fig:distri:settling_fit}.

The fact that the distribution will locally follow a power-law is an important requirement for being able to construct fitting formulas which reproduce the simulated grain size distributions. It allows us \emph{in some cases} to explain the simulation outcomes with the local kernel index (although coagulation and fragmentation are non-local processes in mass space, since each mass may interact with each other mass). A physically motivated recipe to fit the numerically derived distribution functions for the special case of $\xi=11/6$ is presented in Sect.~\ref{sec:distri:recipe}.

\subsection{Non-constant kernels}\label{sec:distri:simresults_nonconstant}
\begin{figure}[t!hb]
  \begin{center}
\makeatletter
\if@referee
    \includegraphics[width=0.7\columnwidth]{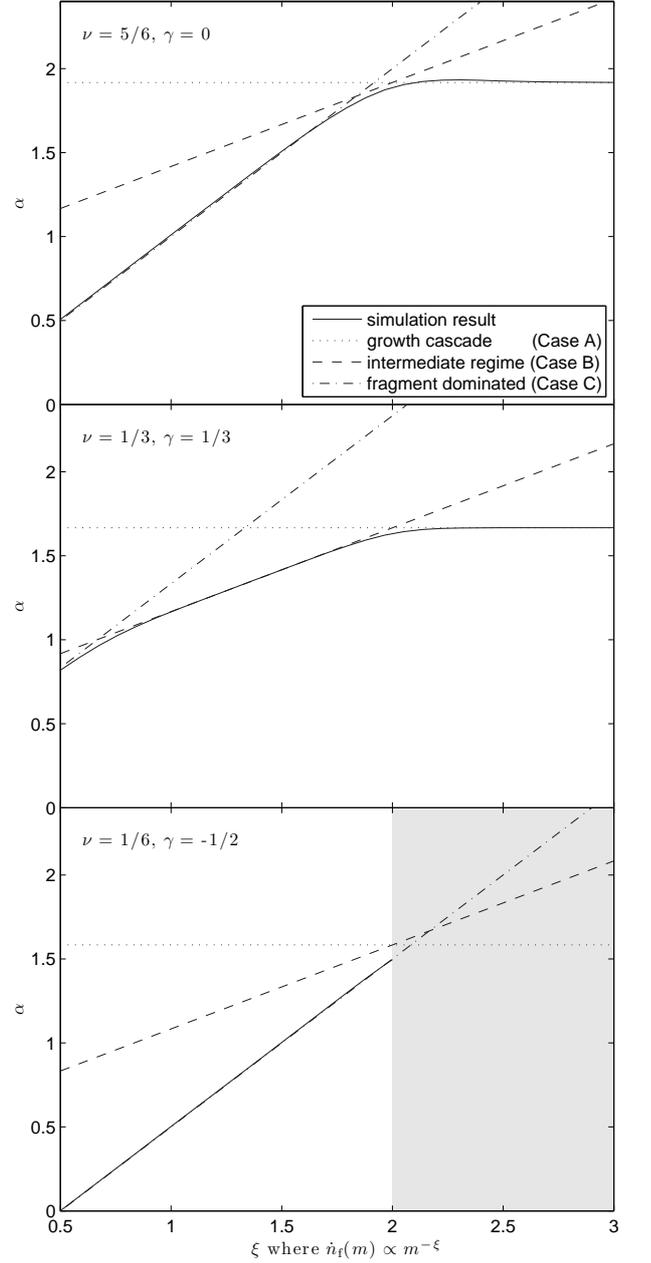}
\else
    \includegraphics[width=0.9\columnwidth]{15228fg5}
\fi
\makeatother
\caption{Exponents of the grain size distributions for three different kernels as function of the fragment distribution index. The plot shows the simulation results (solid lines), the analytic solution for the growth cascade (Case A, dotted lines), the intermediate regime (Case B, dashed lines), and the fragment dominated regime (Case C, dash-dotted lines). The upper panel was calculated with a  $(\nu = 5/6,\gamma=0)$-kernel (i.e. turbulent relative velocities, see Sect.~\ref{sec:distri:diskdistris_vrel}), the middle panel with a $(\nu = 1/3, \gamma=1/3)$-kernel and the lower panel with a $(\nu = 1/6, \gamma=-1/2)$-kernel (i.e. Brownian motion relative velocities). In the grey shaded area, Eq.~\ref{eq:distri:makino_convergence2} is not fulfilled, and the size distribution is not a power-law.}
\label{fig:distri:exponents_both}
\end{center}
\end{figure}

We performed the same tests as in Sect.~\ref{sec:distri:simresults_constant} also for non-constant kernels, i.e. kernels with $\nu>0$. The measured slopes for the cases of $(\nu = 5/6, \gamma = 0)$ (corresponding to the second turbulent regime, see Sect.~\ref{sec:distri:diskdistris_vrel}), $(\nu = 1/3, \gamma = 1/3)$, and $(\nu = 1/6, \gamma = -1/2)$ (i.e., Brownian motion, see Sect.~\ref{sec:distri:diskdistris_vrel}) are shown in Fig.~\ref{fig:distri:exponents_both}. Similar to Fig.~\ref{fig:distri:exponent}, the distribution always follows the minimum index of the three different regimes (Cases A, B, and C). It should be noted that for all indices $\xi$ of the fragmentation law, all processes -- coagulation, fragmentation, and re-distribution of fragments -- take place, but the relative importance of them is what determines the resulting slope.

In Fig.~\ref{fig:distri:exponents_both}, it can be seen that Case B (defined in Sect.~\ref{sec:distri:theory_intermediate}) vanishes for the kernel in the upper panel, while it is present for a large range of $\xi$ for the kernel in the middle panel. This can be explained by the definitions of the three regimes which were summarized in Eq.~\ref{eq:distri:nu_regimes}: with $(\nu = 5/6, \gamma=0)$ (cf. upper panel in Fig.~\ref{fig:distri:exponents_both}), Case B is confined between $\xi=11/6$ and 2. The grain size distribution, therefore, switches almost immediately from being growth dominated (Case A) to fragmentation dominated (Case C). In the case of a kernel with $\nu=1/3$ and $\gamma=1/3$, this range spans from $2/3$ to 2, as can be seen in the central panel of Fig.~\ref{fig:distri:exponents_both}.

The lower panel shows the distribution for a Brownian motion kernel (i.e., $\nu = 1/6$ and $\gamma=-1/2$). The grey shaded area highlights the range of $\xi$ values where our predictions do not apply, that is, Eq.~\ref{eq:distri:makino_convergence2} is no longer fulfilled and thus, the resulting steady-state distributions are no longer power-law distributions.

\section{Grain size distributions in circumstellar disks}\label{sec:distri:diskdistris}
In this section, we will leave the previous ``clean'' kernels and focus on the grain size distribution in circumstellar disks including relative velocities due to Brownian and turbulent motion of the particles, settling effects, and a fragmentation probability as function of particle mass and impact velocity.

Combining all these effects makes it impossible to find simple analytical solutions in the case of a coagulation-fragmentation equilibrium. We will therefore use a coagulation fragmentation code to find the steady-state solutions and to show how the steady-state distributions in circumstellar disks depend on our input parameters.

We will first discuss the model ``ingredients'', i.e. the relative velocities and the prescription for fragmentation/sticking. Afterwards, we will define characteristic sizes at which the shape of the size distribution changes due to the underlying physics. In the last subsection, we will then show the simulation results and discuss the influence of different parameters. 

\subsection{Relative velocities}\label{sec:distri:diskdistris_vrel}
We will now include the effects of relative velocities due to Brownian motion, and due to turbulent mixing.
The Brownian motion relative velocities are given by
\begin{equation}
  \Delta u_\mathrm{BM}=\sqrt{\frac{8 \kb T \;(m_1+m_2)}{\pi \, m_1\, m_2}}
  \label{eq:distri:dv_BM}
\end{equation}
where $\kb$ is the Boltzmann constant and $T$ the mid-plane temperature of the disk.
\citet{Ormel:2007p801} have derived closed form expressions for particle collision velocities induced by turbulence. They also provided easy-to-use approximations for the different particle size regimes which we will use in the following. Small particles (i.e. stopping time of particle $\ll$ eddy crossing time) belong to the first regime of \citet{Ormel:2007p801} with velocities proportional to
\begin{equation}
  \Delta u_\mathrm{I} \propto \left|\St_1-\St_2\right|.
  \label{eq:distri:dv_TM1}
\end{equation}
The relative velocities of particles in the second turbulent regime of \citet{Ormel:2007p801} are given by
\begin{equation}
  \Delta u_\mathrm{II} \propto \sqrt{\St_\mathrm{max}},
  \label{eq:distri:dv_TM2}
\end{equation}
where $\St_\mathrm{max}$ is the larger of the particles Stokes numbers. Velocities in this regime show also a weak dependence on the ratio of the Stokes numbers which we will neglect in the following discussion.

Together with the geometrical cross section $\sigma_\mathrm{geo} = \pi (a_1+a_2)^2$, it is straight-forward to estimate the indices of the kernel, $\nu$ and $\gamma$, as defined in Eq.~\ref{eq:distri:kernel_form} and \ref{eq:distri:gamma_makino} for all these regimes, without settling (assuming that only Brownian motion or turbulent motion dominates the relative velocities). The indices for these three sources of relative particle motion are summarized in Table~\ref{tab:distri:nu_and_gamma} \change{}{(for a derivation, see Appendix~\ref{sec:distri:kernel_index})}.
\begin{table}
  \caption{Kernel indices for the different regimes without settling.}
  \centering
  \begin{tabular}{l|ccl}
Regime&                  $\nu$&         $\gamma$&           upper end\\
\hline
Brownian motion regime&  $\frac{1}{6}$& $-\frac{1}{2}$&     \aBT\\[0.1cm]
Turbulent regime I&      1&             0&                  \aonetwo\\[0.1cm]
Turbulent regime II&     $\frac{5}{6}$& 0&                  \amax\\[0.1cm]
\end{tabular}
\label{tab:distri:nu_and_gamma}
\end{table}
If settling is to be included, then the $\nu$ index for particle sizes above $a_\mathrm{sett}$ has to be increased by $\frac{1}{6}$ (see Sect.~\ref{sec:distri:simresults_settling}) while $\gamma$ remains the same. The $\nu$ and $\gamma$ indices for all three regimes can be found in Table~\ref{tab:distri:nu_and_gamma}.

\begin{figure*}[thb]
  \begin{center}
\includegraphics[width=\hsize]{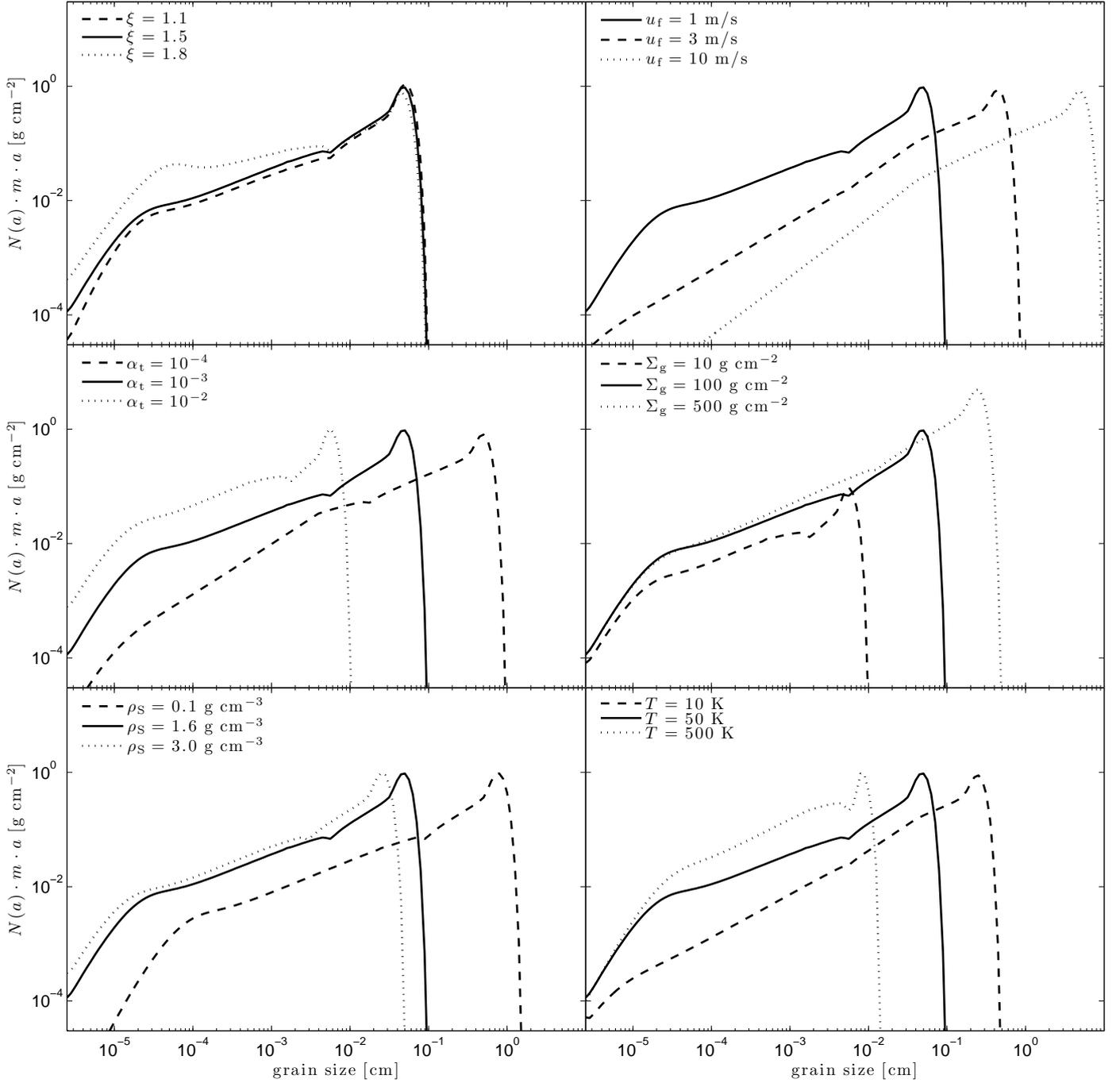}
\caption{Fiducial model and variations of the most important parameters: fragmentation power-law index $\xi$, threshold fragmentation velocity \uf, turbulence parameters \alphat, surface density \Siggas, particle internal density \rhos, and mid-plane temperature $T$. The shape of the vertically integrated size distributions does not depend on the stellar mass or the distance to the star (only via the radial dependence of the parameters above).}
\label{fig:distri:variations}
\end{center}
\end{figure*}

\subsection{Fragmentation and cratering}\label{sec:distri:diskdistris_frag_crater}
We introduce fragmentation and cratering according to the recipe of \citet{Birnstiel:2010p9709}: whether a collision leads to sticking or to fragmentation/cratering is determined by the fragmentation probability
\begin{equation}
p_\text{f} = \left\{
\begin{array}{ll}
0&                              \text{if } \Delta u < \uf - \delta u\\
\\
1&                              \text{if } \Delta u \highlight{\geq} \uf\\
\\
1-\frac{\uf-\Delta u}{\delta u}&    \text{else}
\end{array}
\right.
\end{equation}
which means that all impacts with velocities above the critical break-up threshold \uf lead to fragmentation or cratering while impacts with velocities below $\uf-\delta u$ lead to sticking. The width of the linear transition region $\delta u$ is chosen to be $0.2 \, \uf$, \change{}{since laboratory experiments suggest that there is no sharp fragmentation velocity (\citealp{Blum:1993p4324}; C. G\"uttler 2010, private communication). Our simulation results do not show a strong dependence on this value, which was tested by varying $\delta u$ by a factor of 2.}

The mass ratio of the two particles determines whether the impact completely fragments the larger body (masses within one order of magnitude) or if the smaller particle excavates mass from the larger body (masses differ by more than one order of magnitude). \highlight{This distinction between an erosion and a shattering regime follows the numerical studies of \citet{Paszun:2009p8871} and \citet{Ormel:2009p8002} and experimental studies of \citet{Guttler:2010p9745}. These works do not precisely constrain the mass ratio which distinguishes between both regimes, however our simulation results do only weakly depend on it.}

In the case of fragmentation, the whole mass of both collision partners is redistributed to all masses smaller than the larger body according to Eq.~\ref{eq:distri:frag_powerlaw}.

In the case of cratering, it is assumed that the smaller particle (with mass $m_\mathrm{imp}$) excavates its own mass from the larger body. The mass of the impactor as well as the excavated mass is then redistributed to masses smaller than the impactor mass according to Eq.~\ref{eq:distri:frag_powerlaw}. Thus, the total redistributed mass equals
\begin{equation}
\int_{m_0}^{\highlight{m_\mathrm{imp}}} n(m) \cdot m \, \dx{m} = 2\, m_\mathrm{imp}
\end{equation}
and the mass of the larger body is reduced by $2\, m_\mathrm{imp}$.

Most parameters such as the fragmentation velocity or the amount of excavated material during cratering are not yet well enough constrained (for the most recent experimental work, see \citealp{Blum:2008p1920}, \citealp{Guttler:2010p9745}, and references therein). Experiments suggest fragmentation velocities of a few m s$^{-1}$ and fragment distributions with $\xi$ between 1 and 2 (\citealp{Guttler:2010p9745} find $\xi$ values between 1.07 and 1.37 for SiO$_2$ grains). Simulations of silicate grain growth around 1~AU show that also bouncing (i.e. collisions without sticking or fragmentation) can play an important role \citep[see][]{Weidling:2009p9833,Zsom:2010p9746}. However, changes in material composition such as organic or ice mantles or the monomer size are expected to change this picture. As there is still a large parameter space to be explored, we continue with the rather simple recipe of sticking, fragmentation and cratering outlined above.

\subsection{Regime boundaries}\label{sec:distri:diskdistris_boundaries}
From Eq.~\ref{eq:distri:nu_regimes}, we can calculate the slope of the distribution in the different regimes if we assume that the slope of the distribution at a given grain size always follows from the kernel index $\nu$. To construct a whole distribution consisting of several power-laws for each regime, we need to know where each of the different relative velocity regime applies.

It is important to note that relative velocities due to Brownian motion decrease with particle size whereas the relative velocities induced by turbulent motion increase with particle size (up to $\St=1$). Therefore, Brownian motion dominates the relative velocities for small particles, while for larger particles, turbulence dominates. From numerical simulations, we found that at those sizes where the highest turbulent relative velocities (i.e. collisions with the smallest grains) start to exceed the smallest Brownian motion relative velocities (i.e. collisions with similar sized grains), the slope of the distribution starts to be determined by the turbulent kernel slope. By equating the approximate relative velocity of \citet{Ormel:2007p801} and Eq.~\ref{eq:distri:dv_BM}, the according grain size can be estimated to be
\begin{equation}
a_\mathrm{BT} \approx \left[\frac{8 \Siggas}{\pi \rhos} \cdot \mathrm{Re}^{-\frac{1}{4}} \cdot \sqrt{\frac{\mu \, \mpr}{3\pi \, \alphat}} \cdot \left(\frac{4\pi}{3}\rhos\right)^{-\frac{1}{2}}  \right]^{\frac{2}{5}},
\label{eq:distri:a_01}
\end{equation}
where we approximate the Reynolds number (i.e., the ratio of turbulent viscosity $\nu_\mathrm{t} = \alpha \csound \Hp$ over molecular viscosity) near the disk mid-plane by
\begin{equation}
\Rey \approx \frac{\alphat \, \Siggas \, \sighyd}{2 \, \mu \, \mpr}.
\label{eq:distri:reynolds}
\end{equation}
Here, \sighyd is the cross section of molecular hydrogen (taken to be $2\times~10^{-15}$~cm$^2$) and $\mu=2.3$ is the mean molecular weight in proton masses $\mpr$.

Turbulent relative velocities strongly increase for grains with a stopping time that is larger or about the turn-over time of the smallest eddies. More specifically, the Stokes number of the particles at this change in the relative velocity is
\begin{equation}
\St_\mathrm{12} = \frac{1}{y_a} \, \frac{t_\eta}{t_\mathrm{L}} = \frac{1}{y_a} \, \Rey^{-\frac{1}{2}},
\end{equation}
where $t_\mathrm{L}=1/\Ok$ and $t_\eta = t_\mathrm{L} \cdot \Rey^{-\half}$ are the turn-over times of the largest and the smallest eddies, respectively, and $\Ok$ is the Kepler frequency. \citet{Ormel:2007p801} approximated the factor $y_a$ to be about 1.6. The corresponding grain size in the Epstein regime is therefore given by
\begin{equation}
\aonetwo = \frac{1}{y_a} \, \frac{2 \Siggas}{\pi \, \rhos} \cdot \Rey^{-\frac{1}{2}},
\label{eq:distri:a12}
\end{equation}
As mentioned above, the Brownian motion relative velocities of small grains decrease with their size. For larger sizes, the relative velocities due to turbulent motion are gaining importance, which are increasing with size until a Stokes number of unity. For typical values of the sound speed
\begin{equation}
\csound=\sqrt{\frac{\kb\, T}{\mu \mpr}}
\label{eq:distri:csound}
\end{equation}
and the turbulence parameter $\alphat$, the largest turbulent relative velocity $\Delta u_\mathrm{max} \approx  \sqrt{\alphat}\,\csound$ exceed the critical collision velocity of the grains (which is of the order of a few m s$^{-1}$) and therefore leads to fragmentation of the dust particles. In the case of very quiescent environments and/or larger critical collision velocities, particles do not experience this fragmentation barrier and can continue to grow. Hence, a steady state is never reached.
The work presented here focuses on the former case where
\begin{equation}
\Delta u_\mathrm{max} > \uf,
\label{eq:distri:frag_condition}
\end{equation}
and grain growth is always limited by fragmentation.

As relative turbulent velocities are (in our case) increasing with grain size, we can relate the maximum turbulent relative velocity and the critical collision velocity to derive the approximate maximum grain size which particles can reach \cite[see][]{Birnstiel:2009p7135}
\begin{equation}
a_\mathrm{max} \simeq \frac{2\Siggas }{\pi \alphat \rho_\mathrm{s}} \cdot \frac{\uf^2}{\csound^2}.
\label{eq:distri:a_max}
\end{equation}

\subsection{Resulting steady-state distributions}\label{sec:distri:diskdistris_results}
The  parameter space is too large to even nearly discuss all possible outcomes of steady state grain size distributions. We will therefore focus on a few examples and rather explain the basic features and the most general results only. For this purpose, we will adopt a fiducial model and consider the influences of several parameters on the resulting grain size distribution: $\xi$, \uf, \alphat, \Siggas, \rhos, and $T$ (see Fig.~\ref{fig:distri:variations}).

The fiducial model (see the solid black line in Fig.~\ref{fig:distri:variations}) shows the following features:
steep increase from the smaller sizes until a few tenth of a micrometer. This relates to the regime dominated by Brownian motion relative velocities. The upper end of this regime can be approximated by Eq.~\ref{eq:distri:a_01}. The flatter part of the distribution is caused by a different kernel index $\nu$ in the parts of the distribution which are dominated by turbulent relative velocities. The dip at about 60~$\mu$m (cf. Eq.~\ref{eq:distri:a12}) is due to the jump in relative velocities as the stopping time of particles above this size exceeds the shortest eddy turn-over time \citep[see][]{Ormel:2007p801}.

The upper end of the distribution is approximately at \amax. The increased slope of the distribution and the bump close to the upper end are caused by two processes. Firstly, a boundary effect: grains mostly grow by collisions with similar or larger sized particles. Grains near the upper end of the distribution lack larger sized collision partners and therefore the number density needs to increase in order to keep the flux constant with mass (i.e. in order to keep a steady-state).
Secondly, the bump is caused by cratering: impacts of small grains onto the largest grains do not cause growth or complete destruction of the larger bodies, instead they erode them. Growth of these larger bodies is therefore slowed down and, similar to the former case, the mass distribution needs to increase in order to fulfill the steady-state criterion (``pile-up effect'').

The upper left panel in Fig.~\ref{fig:distri:variations} shows the influence of the distribution of fragments after a collision event: larger values of $\xi$ mean that more of the fragmented mass is redistributed to smaller sizes. Consequently, the mass distribution at smaller sizes increases relative to the values of smaller $\xi$ values.

The strong influence of the fragmentation threshold velocity \uf can be seen in the upper right panel in Fig.~\ref{fig:distri:variations}: according to Eq.~\ref{eq:distri:a_max}, an order of magnitude higher \uf leads to a 100 times larger maximum grain size.

The grain size distributions for different levels of turbulence are shown in the middle left panel of Fig.~\ref{fig:distri:variations}. The effects are two-fold: firstly, an increased \alphat leads to increased turbulent relative velocities, thus, moving the fragmentation barrier \amax to smaller sizes (cf. Eq.~\ref{eq:distri:a_max}).
Secondly, a larger \alphat shifts the transition within the turbulent regime, \aonetwo, to smaller sizes. Consequently, the second turbulent regime gains importance as \alphat is increased since its upper end lower boundary extend ever further.

\begin{figure}[tbh]
  \begin{center}
\includegraphics[width=\hsize]{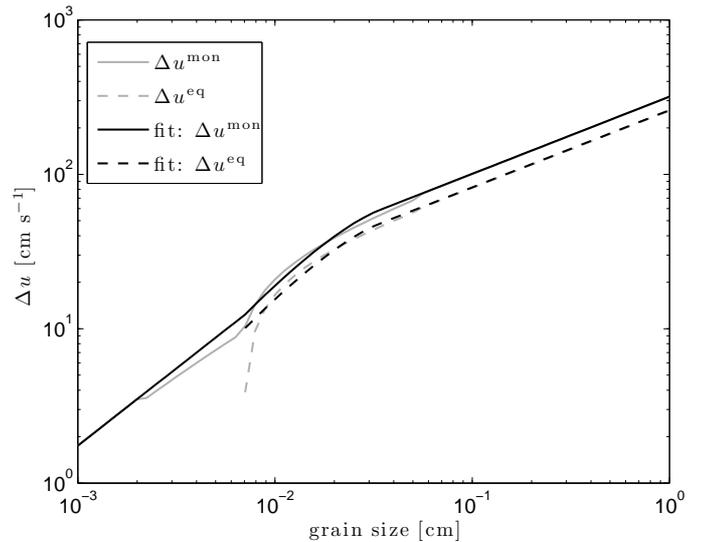}
\caption{Comparison of the turbulent relative velocities of \citet{Ormel:2007p801} to the fitting formula used in the recipe (see Eqs.~\ref{eq:distri:du_eq} and \ref{eq:distri:du_mon}) for grain size distributions. The largest error in the resulting upper grain size $a_\mathrm{P}$ derived from the fitting formula is about 25\%.}
\label{fig:distri:dv_fit}
\end{center}
\end{figure}

The middle right panel in Fig.~\ref{fig:distri:variations} displays the influence of an decreased gas surface density \Siggas (assuming a fixed dust-to-gas ratio). It can be seen that not only the total mass is decreased due to the fixed dust-to-gas ratio but also the upper size of the distribution \amax decreases. This is due to the coupling of the dust to the gas: with larger gas surface density, the dust is better coupled to the gas. This is described by a decreased Stokes number (see Eq.~\ref{eq:distri:st}) which, in turn, leads to smaller relative velocities and hence a larger \amax.
Interestingly, the shape of the grain size distribution does not depend on the total dust mass, but on the total gas mass, as long as gas is dynamically dominating (i.e., $\Siggas \gg \Sigdust$). If more dust were to be present, grains would collide more often, thus, a steady-state would be reached faster and the new size distribution would be a scaled-up version of the former one. However the velocities at which grains collide are determined by the properties of the underlying gas disk. In this way, the dust grain size distribution is not only a measure of the dust properties, but also a measure of the gas disk physics like the gas density and the amount of turbulence.

The shape of the size distribution for different grain volume densities \rhos does not change significantly. However most regime boundaries (\asett, \aonetwo, and \amax) are inversely proportional to \rhos (because of the coupling to the gas, described by the Stokes number). A decrease (increase) in \rhos therefore shifts the whole distribution to larger (smaller) sizes as can be seen in the lower left panel in Fig.~\ref{fig:distri:variations}.

The upper end of the distribution, \amax, is inversely proportional to the mid-plane temperature $T$ (as in the case of the turbulence parameter) whereas the transition between the different turbulent regimes \aonetwo does not. Therefore, increasing the temperature decreases \amax in the same way as decreasing \Siggas does. However \aonetwo is not influenced by temperature changes, therefore the shape of the size distribution changes in a different way than in the case of changing \Siggas as can be seen by comparing the middle right and the  lower right panels of Fig.~\ref{fig:distri:variations}. 
\section{Fitting formula for steady-state distributions}\label{sec:distri:recipe}
In this section, we will describe a simple recipe which allows us to construct vertically integrated grain size distributions which fit reasonably well to the simulation results presented in the previous section.


The recipe does not directly depend on the radial distance to the star or on the stellar mass. A radial dependence only enters via radial changes of the input parameters listed in Table~\ref{tab:distri:glossary}. This recipe has been tested for a large grid of parameter values\footnote{See also 
\href{http://www.mpia.de/distribution-fits}{www.mpia.de/distribution-fits}} (shown in Table~\ref{tab:distri:grid}),
however there are some restrictions.

\begin{figure*}[t!bh]
  \begin{center}
\includegraphics[width=0.75\hsize]{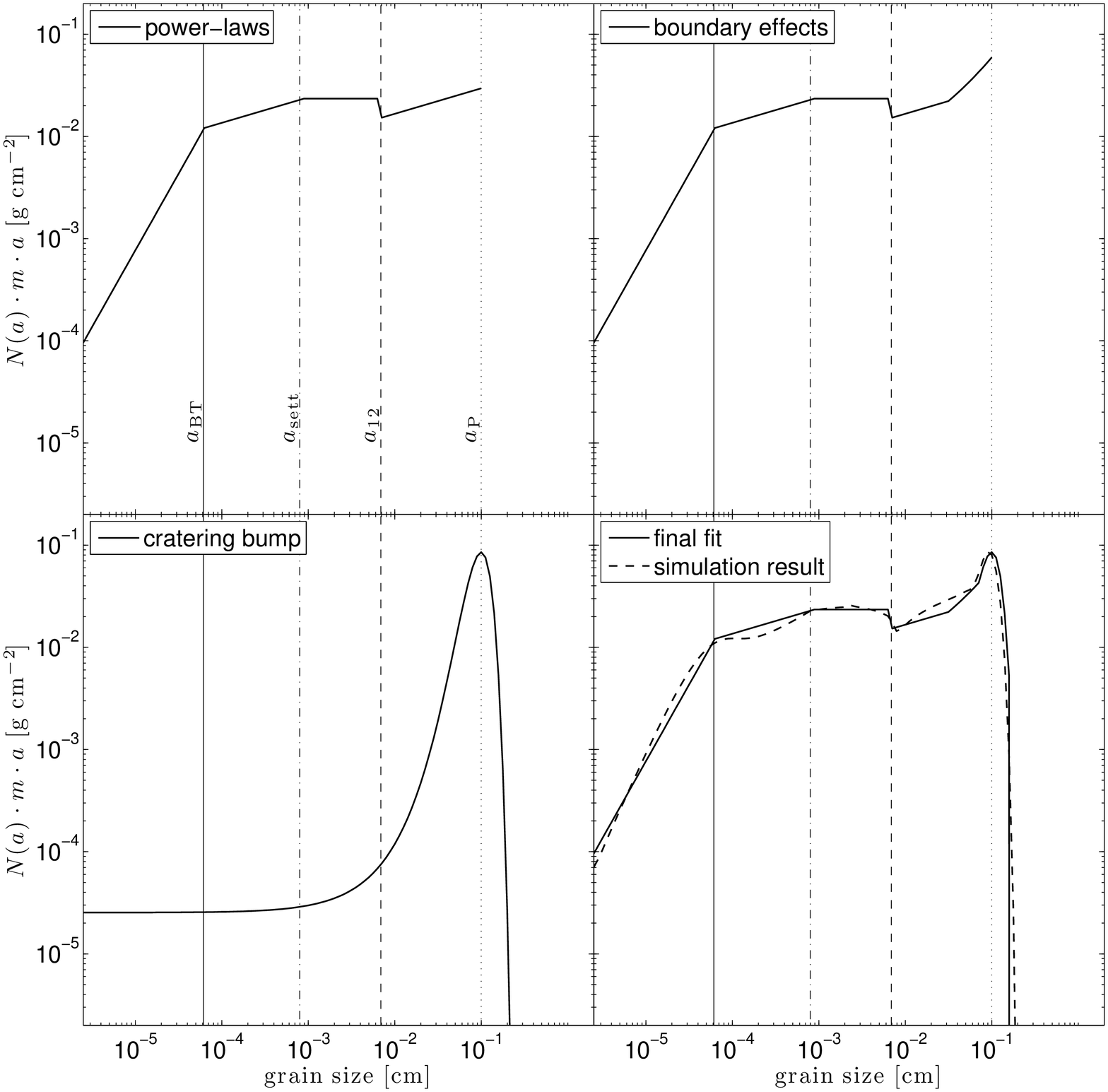}
\caption{Step-by-step construction of the fit distribution for the following parameters: $\Siggas=20$ g cm$^{-2}$, $\Sigdust=0.2$ g cm$^{-2}$, $\alphat=1\e{-4}$, $\uf=1$ m s$^{-1}$, $\xi=1.833$, $T=50$ K, $\rhos = 1.6$ g cm$^{-3}$. \textit{The upper left panel} shows the distribution after step 5: each interval obeys a different power-law, the distribution is continuous apart from a jump at \aonetwo. \textit{The upper right panel} displays the fit including an increase at the upper end, according to step 6. The bump caused by cratering (cf. Eq.~\ref{eq:distri:bump}) is shown in \textit{the bottom left panel} while \textit{the bottom right panel} compares the final fit distribution (solid curve) to the simulation result (dashed curve). The vertical lines correspond to the regime boundaries: \aBT (solid), \asett (dash-dotted), \aonetwo (dashed) and $a_\mathrm{P}$ (dotted).}
\label{fig:distri:app_recipe}
\end{center}
\end{figure*}

\subsection{Limitations}
These fits strictly apply only for the case of $\xi=11/6$. In this case, the slopes of the distribution agree well with the predictions of the intermediate regime (Case B, defined in Sec.~\ref{sec:distri:theory_intermediate}). For smaller values of $\xi$, the slopes do not strictly follow the analytical predictions. This is due to the fact that we include cratering, which is not covered by our theory. Erosion is therefore an important mode of fragmentation: it dominates over complete disruption through the high number of small particles \citep[see also][]{Kobayashi:2010p9774} and it is able to redistribute significant amounts of mass to the smallest particle sizes.

One important restriction for this recipe is the upper size of the particles \amax: it needs to obey the condition of Eq.~\ref{eq:distri:frag_condition}, since otherwise, particles will not experience the fragmenting high-velocity impacts and a steady state will never be reached since particles can grow unhindered over the meter-size barrier.

\begin{table}
\caption{\highlight{Parameter values for which the recipe presented in Sect.~\ref{sec:distri:recipe} has been compared to simulation results: $\alphat$ is the turbulence parameter, $T$ is the mid-plane temperature, \Siggas is the gas surface density, and \uf is the critical collision velocity.}}
\label{tab:distri:grid}
\centering
\highlight{
\begin{tabular}[c]{lrllllll}
\hline\hline
parameter & unit &\multicolumn{6}{c}{values}\\
\hline\\[-0.3cm]
    $\alphat$         &              & $10^{-4}$ & $10^{-3}$ & $10^{-2}$ & -      & -      & -      \\
    $T$               &[K]           & 10        & 100       & 500       & $10^3$ & -      & -      \\
    \Siggas           &[g cm$^{-2}$] & 0.1       & 1         & 10        & 100    & $10^3$ & $10^4$ \\
    \uf               &[m s$^{-1}$]  & 1         & 3         & 10        & -      & -      & -      \\
\hline
\end{tabular}
}
\end{table}

There are also restrictions to a very small \amax: if \amax is close to or even smaller than \aonetwo, then the fit will not represent the true simulation outcome very well. In this case, the upper end of the distribution, and in more extreme cases the whole distribution will look much different. Thus, the sizes should obey the condition
\begin{equation}
5\,\mu\text{m} < \aonetwo < \amax,
\end{equation}
where each inequality should be within a factor of a few.

\subsection{Recipe}
The following recipe calculates the vertically integrated mass distribution of dust grains in a turbulent circumstellar disk within the the above mentioned limitations. The recipe should be applied on a logarithmic grain size grid $a_i$ with a lower size limit of 0.025 $\mu$m and a fine enough size resolution ($a_{i+1}/a_i\lesssim 1.12$). 
For convenience, all variables are summarized in Table~\ref{tab:distri:glossary}.
The steps to be performed are as follows:
\begin{enumerate}
  \item Calculate the grain sizes which represent the regime boundaries \aBT, \aonetwo, \asett which are given by Eqs.~\ref{eq:distri:a_01}, \ref{eq:distri:a12}, and \ref{eq:distri:asett}.
  \item Calculate the turbulent relative velocities for each grain size. For this, we approximate the equations which are given by \citet{Ormel:2007p801}. Collision velocities with monomers are given by
  \begin{equation}
  \Delta u_i^\mathrm{mon} = \ugas\cdot \, \left\{
  \begin{array}{lll}
                    \Rey^\frac{1}{4}\cdot(\St_i-\St_0)       &\mathrm{for}& a_i<\aonetwo\\ \\
    (1-\epsilon)  \cdot\Rey^\frac{1}{4}\cdot(\St_i-\St_0)  &\mathrm{for}& \aonetwo \highlight{\leq} a_i<5\,\aonetwo\\
    \phantom{(1}+ \epsilon\phantom{)} \cdot\sqrt{3\cdot\St_i}\\ \\
    \sqrt{3\cdot\St_i}                                      &\mathrm{for}& a_i \highlight{\geq} 5\,\aonetwo\\ \\
  \end{array}\right.
  \label{eq:distri:du_mon}
  \end{equation}
  where \Rey is the Reynolds number (see Eq.~\ref{eq:distri:reynolds}), and $\St_i$ and $\St_0$ are the Stokes numbers of $a_i$ and monomers \change{}{($a_0 = 0.025 \mu$m)}, respectively (cf. Eq.~\ref{eq:distri:st}). \ugas is given by
  \begin{equation}
    \ugas = \csound \, \sqrt{\frac{3}{2}\alphat},
    \label{eq:distri:ugas}
  \end{equation}
  and the interpolation parameter $\epsilon$ is defined as
  \begin{equation}
    \epsilon = \frac{a_i-\aonetwo}{4\,\aonetwo}.
    \label{eq:distri:rec_epsi}
  \end{equation}
  and collisions with similar sized bodies are approximated as
  \begin{equation}
  \Delta u_i^\mathrm{eq} = \left\{
  \begin{array}{lll}
    0                        &\mathrm{for}& a_i<\aonetwo\\ \\
    \sqrt{\frac{2}{3}}\cdot\Delta u_i^\mathrm{mon}  &\mathrm{for}& a_i>\aonetwo\\
  \end{array}\right.
  \label{eq:distri:du_eq}
  \end{equation}
  A comparison between these approximations and the formulas of \citet{Ormel:2007p801} is shown in Fig.~\ref{fig:distri:dv_fit}.
  \item Using Eqs.~\ref{eq:distri:du_mon} and \ref{eq:distri:du_eq} for the relative velocities, and the transition width $\delta u=0.2\uf$, find the grain sizes which correspond to the following conditions:
  \begin{itemize}
    \item  $a_\mathrm{L}$: particles above this size experience impacts with monomers with velocities of $\Delta u^\mathrm{mon}_i\geq\uf-\delta u$ (i.e., cratering starts to become important).
    \item $a_\mathrm{P}$: particles above this size experience impacts with equal sized grains with velocities of $\Delta u^\mathrm{eq}_i\geq\uf-\delta u$ (i.e., fragmentation becomes important).
    \item $a_\mathrm{R}$: particles above this size experience impacts with similar sized grains with velocities of $\Delta u^\mathrm{eq}_i\geq\uf$ (i.e. every impact causes fragmentation/cratering).
  \end{itemize}
  \item Calculate the factor $J$ according to the recipe
  \begin{align}
    V &= \csound \, \left(\frac{8 \, \mu \mpr \, \Siggas}{\alphat\,\sighyd}\right)^{\frac{1}{4}}
    \sqrt{\frac{3}{4}\, \frac{\alphat}{\Siggas \, \, y_a}} \label{eq:distri:V}\\
    J &= \left(2.5^{-9}+\left(1.1^9+\left(1+2\,\sqrt{3}\, \frac{V}{\uf}\right)^9\right)^{-1}\right)^{-\frac{1}{9}}, \label{eq:distri:J}
  \end{align}
  where $\mu=2.3$, $\sighyd=2\e{-15}$~cm$^2$ and $y_a= 1.6$.
  \item The power-law indices of the mass distribution $\delta_i$ for each interval between the regime boundaries (\aBT, \aonetwo, \asett, $a_\mathrm{P}$) are calculated according to the intermediate regime (cf.~Eq.~\ref{eq:distri:intermediate_slope}).
  The slopes have to be chosen according to the kernel regime (Brownian motion, turbulent regime 1 or 2) and according to whether the regime is influenced by settling or not (i.e., if $a$ is larger or smaller than \asett). The resulting slopes of the mass distribution are given in Table~\ref{tab:distri:slopes}. The first version of the fit $f(a_i)$ (where $a_i$ denotes the numerical grid point of the particles size array) is now constructed by using power-laws ($\propto a_i^{\delta_i}$) in between each of the regimes, up to $a_\mathrm{P}$. The fit should be continuous at all regime boundaries except a drop of 1/J at the transition at \aonetwo. An example of this first version is shown in the top left panel of Fig.~\ref{fig:distri:app_recipe}.
  \item Mimic the cut-off effects which cause an increase in the distribution function close to the upper end: linearly increase the distribution function for all sizes $a_\mathrm{inc} < a < a_\mathrm{P}$: 
  \begin{equation}
    f(a_\mathrm{i}) \rightarrow f(a_\mathrm{i})\cdot \left(2-\frac{a_\mathrm{i}-a_\mathrm{P}}{a_\mathrm{inc}-a_\mathrm{P}}\right),
  \end{equation}
  with
  \begin{equation}
    a_\mathrm{inc} = 0.3\,a_\mathrm{P}.
  \label{eq:distri:a_inc}
  \end{equation}
  The resulting fit after this step is shown in the upper right panel of Fig.~\ref{fig:distri:app_recipe}. 
  \item The bump due to cratering is mimicked by a Gaussian,
  \begin{equation}
    b(a_i) = 2\cdot f(a_\mathrm{L})\cdot \exp\left(-\frac{\left(a_i-a_\mathrm{P}\right)^2}{\sigma^2}\right),
    \label{eq:distri:bump}
  \end{equation}
  where $\sigma$ is defined as
  \begin{equation}
    \sigma = \frac{\min\left(\left| a_\mathrm{R}-a_\mathrm{P}\right|,\left|a_\mathrm{L}-a_\mathrm{P}\right| \right)}{\sqrt{\highlight{\ln}(2)}},
  \end{equation}
  but should be limited to be at least
  \begin{equation}
    \sigma > 0.1\,a_\mathrm{P}.
  \end{equation} 
  \item The fit $\mathcal{F}(a_i)$ is now constructed by using the maximum of $f(a_i)$ and $b(a_i)$ in the following way:
  \begin{equation}
    \mathcal{F}(a_i) = \left\{\begin{array}{lll}
    f(a_i)&                         \mathrm{if }&   a_i \leq a_\mathrm{L}\\
    \\
    \max\left(f(a_i),b(a_i)\right)& \mathrm{if }&   a_\mathrm{L} \leq a_i \leq a_\mathrm{P}\\
    \\
    b(a_i)&                         \mathrm{if }&   a_\mathrm{P} \leq a_i \leq a_\mathrm{R}\\
    \\
    0&                              \mathrm{else}
    \end{array}
    \right.
  \end{equation}
  \item Finally, the fit needs to be normalized to the dust surface density at the given radius. $\mathcal{F}$ is a (yet un-normalized) vertically integrated mass distribution (shown in the bottom right panel of Fig.~\ref{fig:distri:app_recipe}). To translate this mass distribution to a vertically integrated number density distribution $N(a)$, we need to normalize it as 
  \begin{equation}
    N(a) = \frac{\Sigdust}{\int_{a_0}^{\infty}\mathcal{F}(a)\,\dx{\text{ln}a}} \cdot \frac{\mathcal{F}(a)}{m\cdot a}.
\end{equation}
\end{enumerate}

\begin{table}
  \caption{Power-law exponents of the distribution $n(m)\cdot m\cdot a$. The slopes were calculated using the formula for a coagulation/fragmentation equilibrium (Eq.~\ref{eq:distri:intermediate_slope}). Within each regime of relative velocities, it has to be differentiated whether grains are influenced by settling or not.}
  \begin{center}
\begin{tabular}{p{4cm}|c|c}
\hline
\hline
\multirow{2}{*}{Regime} & \multicolumn{2}{c}{$\delta_i$}\\
\cline{2-3}
& $a_i<\asett$&  $a_i\highlight{\geq}\asett$\\
\hline
&&\\[-0.3cm]
Brownian motion regime&  $\frac{3}{2}$& $\frac{5}{4}$\\[0.1cm]
Turbulent regime I&      $\frac{1}{4}$& $0$\\[0.1cm]
Turbulent regime II&     $\frac{1}{2}$& $\frac{1}{4}$\\[0.1cm]
\hline
\end{tabular}
\label{tab:distri:slopes}
\end{center}
\end{table}

\begin{table*}
  \caption{Definition of the variables used in this paper. The variables grouped as ``input variables'' are the parameters of the fitting recipe in Sect.~\ref{sec:distri:recipe}. ``Other variables'' summarizes the definitions of all other variables used in this recipe.}
  \begin{center}
  \setlength{\extrarowheight}{2pt}
\begin{tabularx}{\linewidth}{l|l>{\raggedright\arraybackslash}Xl}
\hline
\hline
\multirow{10}{0.5cm}{
    \rotatebox{90}{
    \large Input variables
    }}
    & \multirow{2}{*}{variable} & \multirow{2}{*}{definition} & \multirow{2}{*}{unit}\\
    \\
\cline{2-4}
    & \alphat &          Turbulence strength parameter &                              -\\
    & \uf &              Fragmentation threshold velocity &                           cm s$^{-1}$\\
    & \Siggas &          Gas surface density &                                        g cm$^{-2}$\\
    & \Sigdust &         Dust surface density &                                       g cm$^{-2}$\\
    & $\xi$ &            Power-law index of the mass 
                         distribution of fragments, see
                         Eq.~\ref{eq:distri:frag_powerlaw} &                          -\\
    & $T$ &              Mid-plane temperature &                                      K\\
    & \rhos &            Volume density of a dust particle&                           g cm$^{-3}$\\
\hline
    \multirow{30}{0cm}{\rotatebox{90}{\large \centering Other variables}}
    &\ccol$a$ &\ccol              grain size, 
                                  $a=\left(3\,m/(4\pi\,\rhos)\right)^{1/3}$ &         cm\\
    & $\alpha$ &                  Slope of the number density
                                  distribution $n(m)\propto m^{-\alpha}$, 
                                  see Eq.~\ref{eq:distri:massdistribution} &          -\\
    & \aBT &                      Eq.~\ref{eq:distri:a_01} &                          cm\\
    & \aonetwo &                  Eq.~\ref{eq:distri:a12} &                           cm\\
    & \amax &                     Eq.~\ref{eq:distri:a_max} &                         cm\\
    & $a_\mathrm{L}$ &            Left boundary of the bump function
                                  $b(a)$, see Sect.~\ref{sec:distri:recipe},
                                  paragraph~3 &                                       cm\\
    & $a_\mathrm{P}$ &            Peak size of the bump function $b(a)$,
                                  see Sect.~\ref{sec:distri:recipe}, paragraph~3 &    cm\\
    & $a_\mathrm{R}$ &            Right boundary of the bump function
                                  $b(a)$, see Sect.~\ref{sec:distri:recipe},
                                  paragraph~3 &                                       cm\\
    & \asett &                    Eq.~\ref{eq:distri:asett} &                         cm\\
    & $a_\mathrm{inc}$ &          Eq.~\ref{eq:distri:a_inc} &                         cm\\
    & $b(a)$ &                    Bump function, see Eq.~\ref{eq:distri:bump} &       -\\
    & $C_{m_1,m_2}$ &             Collision kernel, see
                                  Eq.~\ref{eq:distri:kernel_form} &                   cm$^{3}$ s$^{-1}$\\
    & $\tilde{C}_{m_1,m_2}$ &     Collision kernel including settling effects,
                                  see Eq.~\ref{eq:distri:kernel_form} &               cm$^{2}$ s$^{-1}$\\
    & \csound &                   sound speed, see Eq.~\ref{eq:distri:csound} &       cm s$^{-1}$\\
    & $\delta_i$ &                Slopes of the fit-function,
                                  see Table~\ref{tab:distri:slopes} &                 -\\
    & $\delta u$ &                Width of the transition between sticking
                                  and fragmentation, taken to be 0.2 \uf &            cm s$^{-1}$\\
    &\ccol$\epsilon$ &\ccol       Interpolation parameter, see 
                                  Eq.~\ref{eq:distri:rec_epsi} &                      -\\
    & $\gamma$ &                  Power-law index of the function
                                  $h(m_2/m_1)$ for large ratios of $m_2/m_1$,
                                  as defined in Eq.~\ref{eq:distri:gamma_makino} &    -\\
    &\ccol$J$ &\ccol              Empirical parametrization of the discontinuity,
                                  see Eq.~\ref{eq:distri:J} &                         -\\
    & $K$ &                       Integral defined in
                                  Eq.~\ref{eq:distri:tanaka_flux_2} &                 -\\
    & \kb &                       Boltzmann constant &                                erg K$^{-1}$\\
    &\ccol$m$ &\ccol              mass of the particle, $m=4\pi/3 \,\rhos\,a^3$ &     g\\
    & $m_0$ &                     Monomer mass &                                      g\\ 
    & $m_1$ &                     Mass above which particles fragment &               g\\ 
    & $\mu$ &                     Mean molecular weight in
                                  proton masses, taken to be 2.3 &                    -\\
    & $\mpr$ &                    Proton mass &                                       g\\
    & $n(m)$ &                    Number density distribution &                       g cm$^{-3}$\\
    & $N(m)$ &                    Vertically integrated
                                  number density distribution &                       g cm$^{-2}$\\
    & $\nu$ &                     Degree of homogeneity of the kernel as
                                  defined in Eq.~\ref{eq:distri:kernel_form} &        -\\
    & \Rey &                      Reynolds number, see Eq.~\ref{eq:distri:reynolds} & -\\
    &\ccol\rhos &\ccol            density of a dust grain &                           g cm$^{-3}$\\
    & \St &                       Mid-plane Stokes number in the
                                  Epstein regime, see Eq.~\ref{eq:distri:st} &        -\\
    & \sighyd&                    Cross-section of molecular hydrogen &               cm$^2$\\
    &\ccol$\ugas$ &\ccol          mean square turbulent gas velocity, see 
                                  Eq.~\ref{eq:distri:ugas} &                          cm s$^{-1}$\\
    & $\Delta u_i^\mathrm{mon}$ & Relative velocities between monomers 
                                  and grains of size $a_i$,
                                  see Eq.~\ref{eq:distri:du_mon} &                    cm s$^{-1}$\\
    & $\Delta u_i^\mathrm{eq}$ &  Relative velocities between grains
                                  of size $a_i$, see Eq.~\ref{eq:distri:du_eq} &      cm s$^{-1}$\\
    &\ccol$V$ &\ccol              Parameter used in the calculation of $J$, see 
                                  Eq.~\ref{eq:distri:V} &                             cm s$^{-1}$\\
    &\ccol$y_a$ &\ccol            equals 1.6, parameter from \citet{Ormel:2007p801} & -\\
\hline
\hline
\end{tabularx}
\label{tab:distri:glossary}
\end{center}
\end{table*}

\section{Conclusions}\label{sec:distri:conclusions}
In this work, we generalize the analytical findings of previous works to the case of grain size distributions in a coagulation/fragmentation equilibrium. Under the assumption that all grains above a certain size, \amax, fragment into a power-law distribution $n_\text{f}(m)\propto m^{-\xi}$, we derived analytical steady-state solutions for self-similar kernels and determined three different cases (see \ref{sec:distri:theory_summary}). \change{}{Cratering is not covered by our theory. However our simulations that include cratering agree with the theoretical predictions for a fragmentation law with $\xi=11/6.$}

Results show that dust size distributions in circumstellar disks do not necessarily follow the often adopted MRN power-law distribution of $n(a)\propto a^{-3.5}$ \citep[see][]{Mathis:1977p789,Dohnanyi:1969p7994,Tanaka:1996p2320,Makino:1998p8778,Garaud:2007p405} when both coagulation and fragmentation events operate. We performed detailed simulations of grain growth and fragmentation to test the analytical predictions and found very good agreement between the theory and the simulation results.

We applied the theory to the gaseous environments of circumstellar disks. Unlike the models of \citet{Garaud:2007p405}, the upper end of the size distribution is \change{}{typically} not limited by the growth time scale but by fragmentation because relative velocities increase with grain size and reach values large enough to fragment grains. The shape of the dust distribution is determined by the gaseous environment (e.g., gas surface density, level of turbulence, temperature and others) since the gas is dynamically dominant as long as the gas surface density significantly exceeds the dust surface density. The total dust mass merely provides the normalization of the distribution and the time scale in which an equilibrium is reached. The results presented in this work show that the physics of growth and fragmentation directly link the upper and the lower end of the dust distribution in circumstellar disks.

A ready-to-use recipe for deriving vertically integrated dust size distributions in circumstellar disks for a fixed value of $\xi=11/6$ is presented in Sect.~\ref{sec:distri:recipe}. Although the collision kernel in circumstellar disks is complicated, we found good agreement with our fitting recipe for a fragment distribution with $\xi=11/6$. The recipe can readily be used for further modeling such as disk chemistry or radiative transfer calculations.

\begin{acknowledgements}
We like to thank J\"urgen Blum and Carsten G\"uttler for useful discussions and Antonella Natta, Luca Ricci, Francesco Trotta, Taku Takeuchi, Gijs Mulders, Sean Andrews and the anonymous referee for helpful comments. 
\end{acknowledgements}

\bibliographystyle{aa}
\bibliography{/Users/til/Documents/Papers/bibliography}

\makeatletter
\if@referee
\processdelayedfloats
\pagestyle{plain}
\fi
\makeatother
\appendix
\section{Derivation of the fragment dominated size distribution}\label{sec:distri:appendix}
In this section, we will derive the slope of the size distribution which is dominated by the largest particles. Since in our scenario, the integrals are confined between the monomer mass $m_0$ and the largest particles at the fragmentation barrier \mf, the integrals do not diverge as in the scenario of \citet{Tanaka:1996p2320} and \citet{Makino:1998p8778}, who consider integration bounds of zero and infinity. However, if the first condition of \cite{Makino:1998p8778} is not fulfilled, then the mass flux (cf. Eq.~\ref{eq:distri:tanaka_flux_1}) is dominated by the upper bound of the integral. This is the case which we will consider in the following.

As noted in Sect.~\ref{sec:distri:theory_fragment}, the flux integral (Eq.~\ref{eq:distri:tanaka_flux_1}) can be split into three separate integrals, in order to distinguish cases of $m_2 > m_1$ or $m_2 < m_1$,
\begin{equation}
F(m) \equiv I_1 + I_2 + I_3,
\end{equation}
where $I_1$, $I_2$, and $I_3$ correspond (from left to right) to the three integrals defined in  Eq.~\ref{eq:distri:split_integral}.
We will now evaluate these integrals in the limits of $m_0 \ll m \ll \mf$, using the limiting behavior of $h(m_2/m_1)$ as given in Eq.~\ref{eq:distri:gamma_makino}.

\subsection{First integral}
Carrying out the integration over $m_2$ in $I_1$ leads to
\begin{equation}
\begin{aligned}
I_1 = & \frac{A^2 \cdot h_0}{\nu-\gamma-\alpha+1} \cdot \int_{m_0}^{m/2} \dx{m_1} \, m_1^{\gamma - \alpha +1} \cdot \\
      & \quad   \left[ \mf^{\nu-\gamma-\alpha+1} - \left(m-m_1\right)^{\nu-\gamma-\alpha+1} \right], 
\end{aligned} 
\end{equation}
from which we can derive Eq.~\ref{eq:distri:makino_convergence1}, the first convergence criterion of \citet{Makino:1998p8778}.

We consider the case where this condition does not hold, i.e.,
\begin{equation}
\nu-\gamma-\alpha+1>0.
\end{equation}
Then, the \mf term in brackets dominates over the other term. Thus, the term in brackets is constant and carrying out the integration yields
\begin{equation}
I_1 = \frac{A^2\cdot h_0 \cdot \mf^{\nu-\gamma-\alpha+1}}{(\nu-\gamma-\alpha+1)\cdot (\gamma-\alpha+2)} \cdot \left[\left(\frac{m}{2}\right)^{\gamma-\alpha+2}  - m_0^{\gamma-\alpha+2}\right].
\end{equation}
Now, if the second condition, Eq.~\ref{eq:distri:makino_convergence2} holds, using $m_0 \ll m$, we derive
\begin{equation}
I_1 = \frac{A^2\cdot h_0 \cdot \mf^{\nu-\gamma-\alpha+1}}{(\nu-\gamma-\alpha+1)\cdot (\gamma-\alpha+2)} \cdot \left(\frac{m}{2}\right)^{\gamma-\alpha+2}.
\label{eq:distri:I1}
\end{equation}

\subsection{Second integral}
\begin{figure}[t!bh]
  \begin{center}
\includegraphics[width=\hsize]{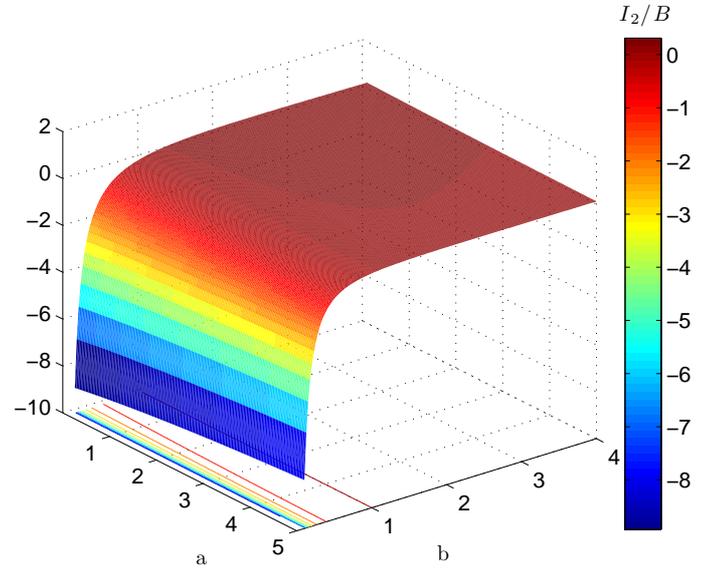}
\caption{Integral $I_2/D$ as function of $a$ and $b$.}
\label{fig:distri:I2}
\end{center}
\end{figure}

We rewrite $I_2$ using the dimensionless variables $x_1 = m_1/m$ and $x_2 = m_2/m$ which yields
\begin{equation}
I_2 = A^2 \cdot h_0 \cdot m^{3+\nu-2\alpha} \int_\frac{1}{2}^1 \dx{x_1} \, \int_{1-x_1}^{x_1} \dx{x_2} \, x_1^{1+\nu-\gamma-\alpha} \cdot x_2^{\gamma-\alpha}.
\end{equation}
By integrating over $x_2$, we derive
\begin{equation}
I_2=\underbrace{\frac{A^2 \cdot h_0\cdot m^{3+\nu-2\alpha}}{\gamma-\alpha+1}}_{:=D} \cdot \int_\frac{1}{2}^{1}\dx{x_1} x_1^{1+\nu-\gamma-\alpha} \cdot \left[x_1^{\gamma-\alpha+1}-\left(1-x_1\right)^{\gamma-\alpha+1} \right].
\end{equation}
The term in square brackets can be split into a sum of integrals, the first of which is straight-forward to evaluate as\balance
\begin{equation}
\frac{I_2}{D} = \frac{1-\left(\frac{1}{2}\right)^{3+\nu-2\alpha}}{\nu-2\alpha+3} -  \int_\frac{1}{2}^{1}\dx{x_1} \, x_1^{\overbrace{1+\nu-\gamma-\alpha}^{a-1}} \cdot \left(1-x_1\right)^{\overbrace{\gamma-\alpha+1}^{b-1}},
\end{equation}
while the second can be identified as a sum of a Beta function $B(a,b)$ and an incomplete Beta function $B_\frac{1}{2}(a,b)$,
\begin{equation}
\frac{I_2}{D} = \frac{1-\left(\frac{1}{2}\right)^{a+b-1}}{a+b-1} -  \left( B(a,b) - B_\frac{1}{2}(a,b) \right).
\end{equation}
The conditions from above,
\begin{alignat}{2}
a&=2+\nu-\gamma-\alpha\quad  &>0 \label{eq:distri:new_condition1} \\
b&=2+\gamma-\alpha    \quad  &>0 \label{eq:distri:new_condition2}
\end{alignat}
assure that the Beta functions, and thus $I_2$, are real. The numerical value of $I_2/D$ for a range of values of $a$ and $b$ are shown in Fig.~\ref{fig:distri:I2}.

\subsection{Third integral}
Similarly, we can rewrite $I_3$ as
\begin{align}
I_3 &= A^2 \cdot h_0 \cdot \int_\frac{1}{2}^m \dx{m_1} \int_{m_1}^{\mf}\dx{m_2}\, m_1^{1+\gamma-\alpha} \cdot m_2^{\nu-\gamma-\alpha}\\
    &= \frac{A^2 \cdot h_0}{\nu-\gamma-\alpha+1} \cdot \int_\frac{m}{2}^m \dx{m_1} \, m_1^{1+\gamma-\alpha} \cdot \left( \mf^{\nu-\gamma-\alpha+1} - \left(\frac{m}{2}\right)^{\nu-\gamma-\alpha+1} \right). \nonumber
\end{align}
If Eq.~\ref{eq:distri:new_condition1} holds, then from $\mf \gg m$ follows
\begin{equation}
I_3 = \frac{A^2 \cdot h_0\cdot \mf^{\nu-\gamma-\alpha+1}}{(\nu-\gamma-\alpha+1)\cdot(\gamma-\alpha+2)} \cdot \left( 1 - \left(\frac{1}{2}\right)^{\gamma-\alpha+2} \right) \cdot m^{\gamma-\alpha+2}.
\end{equation}

\subsection{Deriving the steady-state distribution}
The first and the third integrand $I_1$ and $I_3$ show the same mass dependence and can be summed up to
\begin{equation}
\frac{I_1+I_3}{A^2\cdot h_0} \propto
  \mf^{\nu-\gamma-\alpha+1}\cdot m^{\gamma-\alpha+2}
\label{eq:distri:I13}
\end{equation}
while $I_2$ is proportional to
\begin{equation}
  \begin{aligned}
  \frac{I_2}{A^2 \cdot h_0} &\propto m^{3+\nu-2\alpha} \\
                            &\propto \left(\frac{m}{\mf}\right)^{\nu-\alpha-\gamma+1} \, \mf^{\nu-\gamma-\alpha+1} \,m^{\gamma-\alpha+2} ,
  \end{aligned}
\end{equation}
and therefore
\begin{equation}
\frac{I_2}{I_1+I_3} \propto \left(\frac{m}{\mf}\right)^{\nu-\alpha-\gamma+1}.
\end{equation}
Since the constants of proportionality are factors of order unity and $m<\mf$, the integrals $I_1+I_3$ are much larger than $I_2$, therefore, the flux $F(m)$ is proportional to $m^{\gamma-\alpha+2}$.

In the case of a steady state, this flux and the downward flux of fragments (which is proportional to $m^{2-\xi}$) have to cancel each other. Therefore, the exponents of the mass dependence need to cancel out, i.e.
\begin{equation}
\alpha = \gamma + \xi,
\end{equation}
which is the slope of the steady-state condition in the fragmentation dominated regime (Case C in Fig.~\ref{fig:distri:sketch}).

\section{Derivation of the degree of homogeneity}\label{sec:distri:kernel_index}
\change{}{
In the following, we will derive the degree of homogeneity (cf. Eq.~\ref{eq:distri:kernel_form}) as shown in Table~\ref{tab:distri:nu_and_gamma}. For simplicity, we will drop all constant factors and consider only the proportionalities. The Brownian motion kernel is a product of the geometrical cross section $\sigma_\mathrm{geo} = \pi \, (a_1+a_2)^2$ and the Brownian motion relative velocities (cf. Eq.~\ref{eq:distri:dv_BM}):
\begin{equation}
\begin{split}
C^\mathrm{BM}_{m_1,m_2} &\propto (a_1+a_2)^2 \cdot \sqrt{\frac{m_1+m_2}{m_1\,m_2}}\\
                        &\propto m_1^{\frac{1}{6}} \cdot \underbrace{\left(1+\theta^{\frac{1}{3}} \right)^{2} \, \left(1+\theta^{-1}\right)^{\frac{1}{2}}}_{\equiv h(\theta)},
\end{split}
\end{equation}
where $\theta=m_2/m_1$. Thus, the $m_1$ dependence of the kernel gives $\nu=\frac{1}{6}$. For $\theta\ll 1$, it follows that $h(\theta) \propto \theta^{-\frac{1}{2}}$ and by comparison to Eq.~\ref{eq:distri:gamma_makino}, we can derive $\gamma = -\frac{1}{2}$.
}

\change{}{
The relative velocities in the first regime of turbulent relative velocities is proportional to $\left|a_1-a_2\right|$ (cf. Eq.~\ref{eq:distri:dv_TM1}), thus, the kernel can be written as
\begin{equation}
\begin{split}
C^\mathrm{I}_{m_1,m_2} &\propto (a_1+a_2)^2 \cdot \left|a_1-a_2\right|\\
                        &\propto m_1 \cdot \underbrace{(1+\theta^{\frac{1}{3}})^2\cdot \left| 1-\theta^{\frac{1}{3}}\right|}_{\equiv h(\theta)}.
\end{split}
\end{equation}
In this case, the function $h(\theta)$ is different: $h(\theta)\propto 1$ for $\theta\ll 1$, and we derive $\nu=1$ and $\gamma=0$.
}

\change{}{
In the second turbulent regime, the relative velocities are proportional to the square root of the Stokes number of the larger particle (see Eq.~\ref{eq:distri:dv_TM2}). By using a limit representation of $\max(a_1,a_2)$, we can write
\begin{equation}
\begin{split}
C^\mathrm{II}_{m_1,m_2} &\propto (a_1+a_2)^2 \cdot \lim_{N\rightarrow\infty} \sqrt{(a_1^N+a_2^N)^{\frac{1}{N}}}\\
                        &\propto m_1^\frac{5}{6} \cdot \underbrace{\left(1+\theta^\frac{1}{3}\right)^2 \cdot \lim_{N\rightarrow\infty} (1+\theta^\frac{N}{3})^\frac{1}{2\,N}}_{\equiv h(\theta)}.
\end{split}
\end{equation}
Thus, for the limit of small $\theta$, we derive $\nu=\frac{5}{6}$ and $\gamma = 0$. If settling is included (see Eq.~\ref{eq:distri:settling_c}), then $\nu$ increases by an additional factor of 1/6.
}

\section{Estimating the equilibration time scale}\label{sec:distri:timescale}
\highlight{
As shown in the appendix of \citet{Birnstiel:2010p9709}, monodisperse growth (i.e. assuming all particles have the same size) provides a good estimate of growth time scales. In this approximation, the growth rate is given by
\begin{equation}
\frac{\mathrm{d} a}{\mathrm{d} t} = \frac{1}{4\pi\,\rhos\,a^2} \cdot \frac{\mathrm{d} m}{\mathrm{d} t} \simeq \frac{1}{4\pi\,\rhos\,a^2} \cdot \frac{m}{\tau_\mathrm{coll}},
\label{eq:distri:growthrate}
\end{equation} 
where $\tau_\mathrm{coll}$ is the collision time scale. Integration of Eq.~\ref{eq:distri:growthrate} yields the time a particle needs to grow from size $a_1$ to $a_2$. For Brownian motion and the first two turbulent velocity regimes (see Eqs.~\ref{eq:distri:dv_BM} and \ref{eq:distri:dv_TM2}), we can derive the growth times
\begin{align}
t_\mathrm{BM}(a_1,a_2) &= \frac{1}{5\,\Sigdust\,\Ok} \, \sqrt{\frac{2\,(\pi\rhos)^3}{3\, \mu \mpr}} \, \left(a_2^\frac{5}{2}-a_1^\frac{5}{2}\right)
\label{eq:distri:tau_BM}\\
t_\mathrm{IIa}(a_1,a_2) &= \frac{4}{\Sigdust\,\Ok} \cdot \sqrt{\frac{\Siggas\,\rhos}{3 \alphat}} \, \left(\sqrt{a_2}-\sqrt{a_1}\right) 
\label{eq:distri:tau_TM2a}\\
t_\mathrm{IIb}(a_1,a_2) &= \frac{\Siggas}{\Sigdust\,\Ok\,} \cdot \sqrt{\frac{8}{3\pi}} \cdot \ln\left(\frac{a_2}{a_1}\right). 
\label{eq:distri:tau_TM2b}
\end{align}
Here the last two growth times belong to two distinct cases: $t_\mathrm{IIa}$ is the time in the second turbulent regime (see Eq.~\ref{eq:distri:dv_TM2}) without settling effects (i.e., $a_2<\asett$) while $t_\mathrm{IIb}$ includes the fact that settling of particles to the mid-plane increases the dust density at the mid-plane (denoted by $\rhodust$) and thus accelerates growth.
}

\highlight{
Typically, coagulation starts by Brownian motion growth from sub-$\mu$m sized particles to sizes where turbulent velocities become important (i.e., sizes larger than $\aBT$, see Eq.~\ref{eq:distri:a_01}). Therefore, we can estimate this time by
\begin{equation}
\tau_\mathrm{BM} = t_\mathrm{BM}(a_0,\aBT),
\end{equation}
where $a_0$ is the monomer size.
}

\highlight{
We will neglect the time needed by particles to grow until the second turbulent regime because it is typically much shorter than the other time scales involved. If particles in the second turbulent regime are already influenced by settling ($a_1>\asett$), then growth proceeds according to Eq.~\ref{eq:distri:tau_TM2b} and the time is given by
\begin{equation}
\tau_\mathrm{II} = t_\mathrm{IIb}(\aonetwo,\amax).
\end{equation}
If the size range is entirely below \asett, then the timescale is given by
\begin{equation}
\tau_\mathrm{II}=t_\mathrm{IIa}(\aonetwo,\amax).
\end{equation}
For the case that $\aonetwo\leq\asett\leq\amax$, we need to add both contributions
\begin{equation}
\tau_\mathrm{II} = t_\mathrm{IIa}(\aonetwo,\asett) + t_\mathrm{IIb}(\asett,\amax).
\end{equation}
By comparison to our time evolving simulations of particle growth and fragmentation, we found that
\begin{equation}
\tau_\mathrm{equil} = 8 \cdot\left(\tau_\mathrm{BM} + \tau_\mathrm{II}\right)
\end{equation}
estimates within a factor of a few the time at which the distribution reaches a steady state.
}
 
\end{document}